\documentclass[a4paper,fleqn]{cas-sc}

\usepackage[authoryear,longnamesfirst]{natbib}
\usepackage{pdflscape}
\usepackage{afterpage}
\usepackage{capt-of}
\def\tsc#1{\csdef{#1}{\textsc{\lowercase{#1}}\xspace}}
\tsc{WGM}
\tsc{QE}
\tsc{EP}
\tsc{PMS}
\tsc{BEC}
\tsc{DE}
\begin{document}
\let\WriteBookmarks\relax
\def\floatpagepagefraction{1}
\def\textpagefraction{.001}

% Short title
\shorttitle{Towards FATE in AI for Social Media and Healthcare: A Systematic Review}

% Short author
\shortauthors{Aditya Singhal et~al.}

% Main title of the paper
\title [mode = title]{Towards FATE in AI for Social Media and Healthcare: A Systematic Review}                      
% Title footnote mark
% eg: \tnotemark[1]
% \tnotemark[1,2]

% Title footnote 1.
% eg: \tnotetext[1]{Title footnote text}
% \tnotetext[<tnote number>]{<tnote text>} 
% \tnotetext[1]{This document is the results of the research
%    project funded by the National Science Foundation.}

% \tnotetext[2]{The second title footnote which is a longer text matter
%    to fill through the whole text width and overflow into
   % another line in the footnotes area of the first page.}

% First author
%
% Options: Use if required
% eg: \author[1,3]{Author Name}[type=editor,
%       style=chinese,
%       auid=000,
%       bioid=1,
%       prefix=Sir,
%       orcid=0000-0000-0000-0000,
%       facebook=<facebook id>,
%       twitter=<twitter id>,
%       linkedin=<linkedin id>,
%       gplus=<gplus id>]
\author[1]{Aditya Singhal}[
                        orcid=0000-0001-9634-4075]

% Corresponding author indication
\cormark[1]

% Footnote of the first author
% \fnmark[1]

% Email id of the first author
\ead{asinghal@lakeheadu.ca}

% URL of the first author
% \ead[url]{www.cvr.cc, cvr@sayahna.org}

%  Credit authorship
\credit{Conceptualization of this study, Methodology, Software, Writing - Original draft preparation}

% Address/affiliation
\affiliation[1]{organization={Lakehead University},
    addressline={955 Oliver Rd.}, 
    city={Thunder Bay},
    % citysep={}, % Uncomment if no comma needed between city and postcode
    postcode={P7B 3S3}, 
    % state={},
    country={Canada}}

% Second author
\author[2]{Hasnaat Tanveer}[orcid = 0009-0001-0171-7957]

% Email id of the first author
\ead{hasnaat1278@gmail.com}

% URL of the first author
% \ead[url]{www.cvr.cc, cvr@sayahna.org}

%  Credit authorship
\credit{Methodology}

% Address/affiliation
\affiliation[2]{organization={Superior Collegiate Vocational Institute},
    addressline={333 High St N}, 
    city={Thunder Bay},
    % citysep={}, % Uncomment if no comma needed between city and postcode
    postcode={P7A 5S3}, 
    % state={},
    country={Canada}}
    
% Third author
\author[1]{Vijay Mago}[%
  orcid=0000-0002-9741-3463
   ]
% \fnmark[2]
\ead{vmago@lakeheadu.ca}
% \ead[URL]{www.sayahna.org}

\credit{Supervision}

% Address/affiliation
% \affiliation[2]{organization={Sayahna Foundation},
%     % addressline={}, 
%     city={Jagathy},
%     % citysep={}, % Uncomment if no comma needed between city and postcode
%     postcode={695014}, 
%     state={Trivandrum},
%     country={India}}

% Fourth author
% \author%
% [1,3]
% {Rishi T.}
% \cormark[2]
% \fnmark[1,3]
% \ead{rishi@stmdocs.in}
% \ead[URL]{www.stmdocs.in}

% \affiliation[3]{organization={STM Document Engineering Pvt Ltd.},
%     addressline={Mepukada}, 
%     city={Malayinkil},
%     % citysep={}, % Uncomment if no comma needed between city and postcode
%     postcode={695571}, 
%     state={Trivandrum},
%     country={India}}

% Corresponding author text
\cortext[cor1]{Corresponding author}
% \cortext[cor2]{Principal corresponding author}

% Footnote text
% \fntext[fn1]{This is the first author footnote. but is common to third
%   author as well.}
% \fntext[fn2]{Another author footnote, this is a very long footnote and
%   it should be a really long footnote. But this footnote is not yet
%   sufficiently long enough to make two lines of footnote text.}

% % For a title note without a number/mark
% \nonumnote{This note has no numbers. In this work we demonstrate $a_b$
%   the formation Y\_1 of a new type of polariton on the interface
%   between a cuprous oxide slab and a polystyrene micro-sphere placed
%   on the slab.
%   }

% Here goes the abstract
\begin{abstract}
As artificial intelligence (AI) systems become more prevalent, ensuring fairness in their design becomes increasingly important. This survey focuses on the subdomains of social media and healthcare, examining the concepts of fairness, accountability, transparency, and ethics (FATE) within the context of AI. We explore existing research on FATE in AI, highlighting the benefits and limitations of current solutions, and provide future research directions. We found that statistical and intersectional fairness can support fairness in healthcare on social media platforms, and transparency in AI is essential for accountability. While solutions like simulation, data analytics, and automated systems are widely used, their effectiveness can vary, and keeping up-to-date with the latest research is crucial.

\end{abstract}

% Use if graphical abstract is present
% \begin{graphicalabstract}
% \includegraphics{figs/grabs.pdf}
% \end{graphicalabstract}

% Research highlights
% \begin{highlights}
% \item We provide a thorough examination of the concepts of fairness, accountability, transparency, and ethics in the context of AI systems, with a particular focus on social media and healthcare domains.

% \item We present an extensive analysis of existing research on FATE in AI, highlighting the benefits and limitations of current solutions. We discuss various computational methods and approaches employed in achieving fairness and accountability in AI systems, and we explore the evaluation metrics used to assess their effectiveness.

% \item We identify gaps and challenges in the field and propose future research directions to advance the understanding and implementation of FATE in AI for social media and healthcare applications. Our recommendations aim to promote the development of ethically responsible AI systems that prioritize fairness and transparency.
% \end{highlights}

% Keywords
% Each keyword is seperated by \sep
\begin{keywords}
fairness \sep accountability \sep transparency \sep ethics \sep artificial intelligence \sep social media \sep healthcare
\end{keywords}

\maketitle

\section{Introduction}
\subsection{Background}

Machine learning algorithms are utilized by all stakeholders in today’s world. Most fields, from governance to financial decision-making, and medical diagnosis to security assessment, depend on artificial intelligence (AI) to deliver results. On the surface, this progression towards automation has clear benefits: it is fast and reliable, and cost-effective for businesses over time \cite{mehrabi2021survey}. However, as research in AI continues to advance at a rapid pace, it is becoming increasingly important to ensure that its development and deployment are guided by principles of fairness, accountability, transparency, and ethics.

The vast amount of user data available on social media platforms (SMPs) can be used to identify patterns, trends, and behaviour. SMPs such as Twitter are predominantly used by young and urban residents \cite{mashhadi2021no}. They also have minimum age requirements, leading any machine learning algorithm trained on data from these sources to be biased toward a certain demographic. The wide availability of social media data is also opportunistic for health-related research \cite{leonelli2021fair}. However, lack of ethical oversight at the data collection stage of a research project could lead to the inclusion of data from users who did not consent to it, thereby raising questions on \textit{‘who is a participant of the study?’}. The content on SMPs is also heavily influenced by \textit{‘social’} processes and can not be taken at its face value. Certain topics might generate traction from users of particular areas or demographics \cite{singhal2022synergy}, and the trustworthiness of data continues to be a challenge \cite{leonelli2021fair}.  The lack of availability of code behind machine learning algorithms in propriety software makes it difficult to analyze the underlying patterns which may cause discriminative decisions.

\textit{Misinformation} refers to the inadvertent dissemination of false information on a topic, while \textit{disinformation} is the intentional spread of false information for motives such as financial gain, fame, or damaging the reputation of others \cite{kington2021identifying}. The proliferation of false information is prevalent on social media, and during the COVID-19 pandemic, there was widespread misinformation about vaccines, including unfounded claims that they were harmful. Such misinformation led to doubts about the government and vaccine hesitancy, posing risks to public health and efforts to control the spread of COVID-19. To counter this issue, AI is being employed to help identify and label reliable and high-quality information for users. Reliable AI systems are trained using accurate health information from reputable sources such as non-profit health organizations or government agencies, ensuring that the information provided is based on sound scientific evidence. In addition, corporations are advised to prioritize transparency to build trust among the public and foster a better community. Transparent practices can enhance accountability and credibility, leading to increased trust from the people and promoting a positive environment of mutual trust and understanding.

On the positive side, social media can serve as a platform for users to share new health information, allowing the health sector to potentially access more medical insights and knowledge \cite{pirraglia2013social}. However, the drawbacks of social media, such as the lack of verifiability and potential misinformation, need to be carefully addressed to ensure accurate and reliable health information dissemination.
The FATE research focuses on evaluating the fairness and transparency of AI models, developing metrics to assess the accountability of AI systems, and designing frameworks for responsible and ethical AI development. The ethical implications of using algorithms and AI are closely dependent on the practices of transparency and accountability \cite{wieringa2020account}. Algorithmic systems can be viewed as socio-technical systems that are involved in many different sectors, such as culture, programming, laws and more. One way to avoid discrimination is to have a human in the loop of these algorithmic processes when needed. For example, when the US judicial system COMPAS decides on the likelihood of a prisoner committing another crime after leaving prison, a judge should review the decision of the AI first in order to check the accuracy of the decision. Since the current AI systems are expected to have some bias in their decision models, ensuring that researchers are implementing these models in an organized, ethical, and systematic manner, would make it easier to enforce accountability of actions \cite{hutchinson2021towards}. Efforts are being made by computer scientists to make AI more transparent by revealing the decision-making process that leads to the final AI answer \cite{johnson2019ai}. This helps in identifying problems or biases and holding individuals accountable in case of failures. The European Union has recommended seven key principles to ensure ethical AI, including human agency and oversight, technical robustness and safety, privacy and data governance, transparency, diversity, nondiscrimination and fairness, social and environmental well-being, and accountability.

\subsection{Motivation}

Studies on the FATE of AI in social media surveillance have focused on ensuring that AI systems used for monitoring online content and activity do not perpetuate existing biases or discrimination. For example, research has shown that algorithms used in social media surveillance may have biases against certain groups, or that algorithms used to detect hate speech may not be effective in detecting it against all groups. Recent research has also mainly focused on one aspect of understanding machine learning models. The research in AI ethics is heavily influenced by geographic locations and socio-economic factors \cite{hagerty2019global}. There have been several discussions on the best practices for evaluating work produced by explanatory AI (XAI) and gap analyses performed on model interpretability in AI \cite{gilpin2018explaining}, \cite{chakraborty2017interpretability}. The latest developments in machine learning interpretability have also been reviewed previously \cite{carvalho2019machine}. Table \ref{table1} provides an overview of existing review studies discussing FATE in different forms. Therefore, the motivation behind this survey is to present a comprehensive overview of the various computational methods for FATE and provide direction for future research in the field.

\begin{table}[]

\caption{An overview of existing survey articles focusing on FATE. \textit{A = Definitions, B = Computational methods and approaches, C = Evaluation metrics}}
\label{table1}
\begin{tabular}{L|LLL|LLL|LLL|LLL}

\hline
\textbf{Paper}                                                                                                                                                                              & \multicolumn{3}{l}{\textbf{Fairness}}                                                              & \multicolumn{3}{l}{\textbf{Accountability}}                                                        & \multicolumn{3}{l}{\textbf{Transparency}}                                                          & \multicolumn{3}{l}{\textbf{Ethics}}                                                                \\ \hline
                                                                                                                                                                                                 & \textbf{A \hspace{4pt}} & \textbf{B \hspace{4pt}} & \textbf{C} & \textbf{A \hspace{12pt}} & \textbf{B  \hspace{12pt}} & \textbf{C} & \textbf{A  \hspace{12pt}} & \textbf{B  \hspace{12pt}} & \textbf{C} & \textbf{A  \hspace{8pt}} & \textbf{B  \hspace{8pt}} & \textbf{C} \\
\cite{mehrabi2021attributing}                                                                                                                                                & \checkmark                   & \checkmark                                            & \checkmark                          &                      &                                               &                             &                      &                                               &                             &                      &                                               &                             \\
\cite{golder2017attitudes}                                                                                                                 &                      &                                               &                             &                      &                                               &                             &                      &                                               &                             &                      & \checkmark                                            & \checkmark                          \\
\cite{bear2022scoping}                &                      & \checkmark                                            & \checkmark                          & \checkmark                   &                                               &                             &                      &                                               &                             &                      &                                               &                             \\
\cite{attard2022ethics}                                                                                                                         & \checkmark                   & \checkmark                                            & \checkmark                          & \checkmark                   & \checkmark                                            & \checkmark                          & \checkmark                   & \checkmark                                            & \checkmark                          & \checkmark                   &                                               & \checkmark                          \\
\cite{wieringa2020account}                                                                                                                                              &                      &                                               &                             & \checkmark                   & \checkmark                                            & \checkmark                          &                      &                                               &                             &                      &                                               &                             \\
\cite{adadi2018peeking}                                                                                                                    &                      &                                               &                             &                      &                                               &                             &                      & \checkmark                                            & \checkmark                          &                      &                                               &                             \\
\cite{bose2005lineage}                                                                                                                                       &                      &                                               &                             &                      &                                               &                             & \checkmark                   &                                               &                             &                      &                                               &                             \\
\cite{carvalho2019machine}                                                                                                                              & \checkmark                   &                                               &                             &                      &                                               & \checkmark                          & \checkmark                   & \checkmark                                            & \checkmark                          &                      &                                               & \checkmark                          \\
\cite{chakraborty2017interpretability}                                                                                                                                    & \checkmark                   &                                               &                             & \checkmark                   &                                               &                             & \checkmark                   &                                               & \checkmark                          &                      &                                               &                             \\
\cite{hagerty2019global}                                                                                             &                      &                                               &                             &                      &                                               &                             &                      &                                               &                             & \checkmark                   &                                               & \checkmark                          \\
\cite{kalkman2019responsible}                                                                                           &                      &                                               &                             &                      &                                               & \checkmark                          &                      &                                               & \checkmark                          & \checkmark                   & \checkmark                                            & \checkmark                          \\
\cite{world2016medicines}                                                           &                      &                                               &                             &                      &                                               & \checkmark                          &                      &                                               & \checkmark                          &                      &                                               &                             \\
Our paper                                                                                                                                                                                        & \checkmark                   & \checkmark                                            & \checkmark                          & \checkmark                   & \checkmark                                            & \checkmark                          & \checkmark                   & \checkmark                                            & \checkmark                          & \checkmark                   & \checkmark                                            & \checkmark                          \\ \hline
\end{tabular}
\end{table}

We aim to address the following research questions in this work:

\textbf{RQ1}: What are the existing solutions to FATE (Fairness, Accountability, Transparency, and Ethics) when discussing healthcare on Social Media Platforms (SMPs)?

\textbf{RQ2}: How do the different solutions identified in response to RQ1 compare to each other in terms of computational methods, approaches, and evaluation metrics?

\textbf{RQ3}: What is the strength of evidence supporting the different solutions?

The objective of this research is to identify gaps in the current literature and complement existing work in the FATE space by showcasing how various techniques, themes, and contextual considerations can be combined to support social media interventions in healthcare settings.

\subsection{Research Methodology}
Our research methodology is based on the approach presented by the authors of \cite{kofod2012structured}. We utilized Google Scholar\footnote{https://scholar.google.com}, the largest repository of scholarly articles, to perform a strategic search using Table \ref{table2} as a filter to identify research papers relevant to our study. Each of the groups in the table can be customized to retrieve different sets of literature, with the aim of finding the intersection of these sets. This search strategy involves using the AND and OR operators, where the OR operator can be used within the groups and the AND operator between the groups.

\begin{table}[H]
\caption{Search strategy for finding research articles. \textit{T = Search term, G = Group, Quality = \{fairness, accountability, transparency, ethics\}}}
\label{table2}
\begin{tabular}{llll}
\hline
\textbf{}   & \textbf{G1} & \textbf{G2}                 & G3           \\ \hline
\textbf{T1} & Quality & Natural language processing & Social media \\
\textbf{T2} &             & Artificial intelligence     & Healthcare  \\
\textbf{T3} &             & Computer science            &              \\ \hline
\end{tabular}
\end{table}

The search strategy employed for this study can be summarized as follows: (T1G1 AND T1G2) AND (T1G1 AND T2G2) AND (T1G1 AND T3G2) OR (T1G1 AND T1G3) AND (T1G1 AND T2G3). Initially, this search yielded a substantial number of results, which were then filtered using the following steps: (1) considering articles published after 2012, (2) including articles with a generally high citation score (>100), with some exceptions for recent articles with citations < 100, and (3) removing all duplicate articles. Subsequently, a quality assessment of all articles was conducted based on inclusion (IC) and quality criteria (QC). The inclusion criteria consisted of IC1 (the study's main concern is FATE while discussing healthcare on SMPs), IC2 (the study is a primary study presenting empirical results), and IC3 (the study focuses on definitions, computational methods, approaches, and evaluation metrics). The quality criteria included QC1 (clear statement of the research aim). These criteria were applied through a three-stage process: abstract inclusion criteria screening, full-text inclusion criteria screening, and full-text quality screening. This process helped us determine the relevance and quality of the articles for our study.

\begin{figure}[h!]
% \begin{adjustwidth}{-\extralength}{0cm}
\centering
\includegraphics[width=0.5\textwidth]{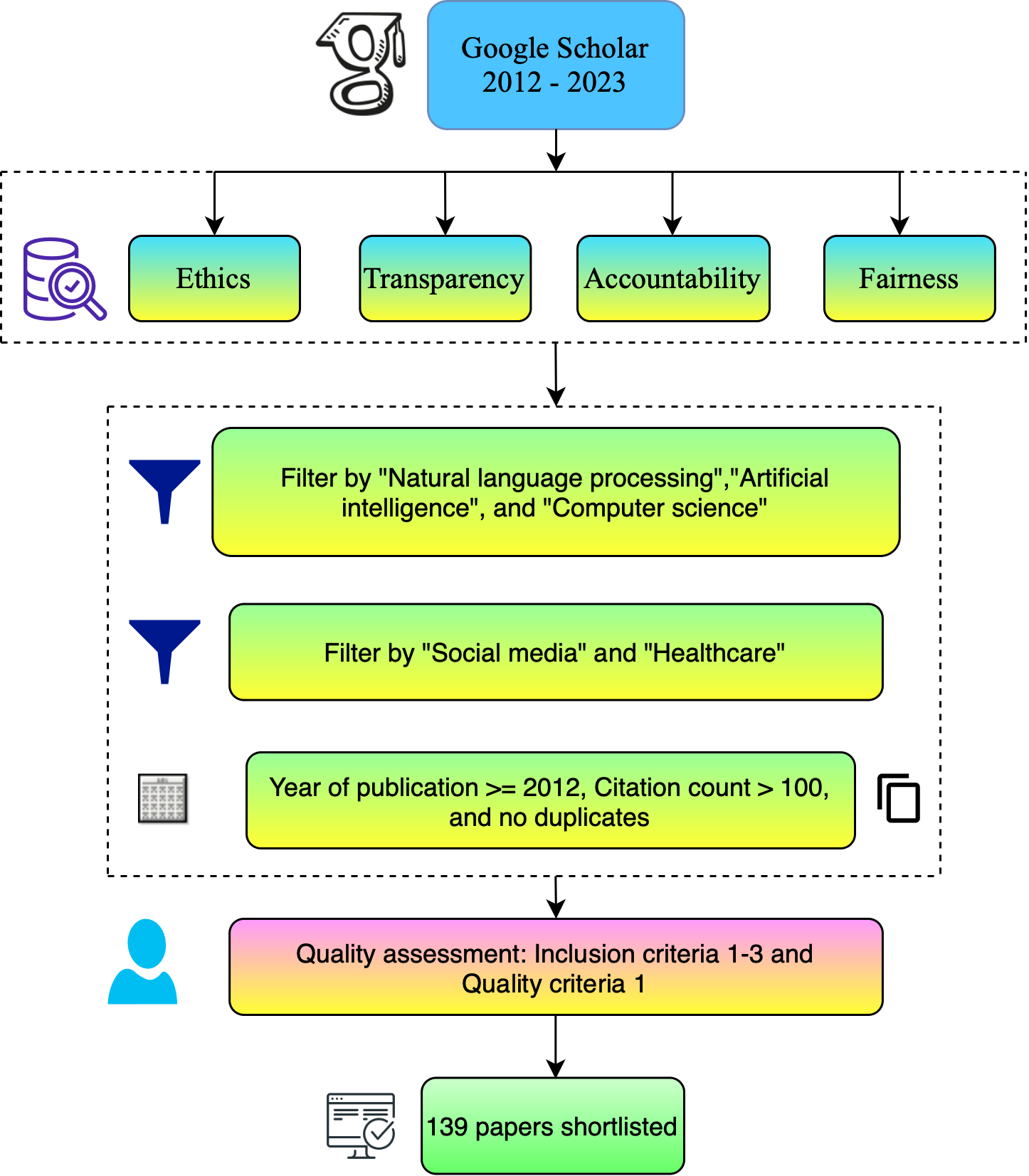}
% \end{adjustwidth}
\caption{Overview of the search strategy and research methodology.\label{fig1}}
\end{figure}

% \section{Definitions, Computational methods and Approaches}

Figure \ref{fig1} outlines the structure and overall, this survey provides an in-depth understanding of one of the most socially important problems in AI for new researchers. The article is structured as follows: Sections \ref{fairness}, \ref{accountability}, \ref{transparency}, and \ref{ethics} cover the definitions, computational methods, approaches, and evaluation metrics for FATE in AI. Section \ref{datasets} provides an overview of FATE in datasets, while section \ref{discussion} offers a discussion on the topic. Finally, future research directions and conclusions are presented in sections \ref{future} and \ref{conclusion}, respectively.

\section{An Overview of Fairness}\label{fairness}

\subsection{Definitions, Computational Methods, and Approaches to Fairness}
The extent to which the general public understands the definition of fairness varies \cite{saha2020measuring}. There are several different definitions that have been proposed in the context of artificial intelligence. 

\subsubsection{{\textbf{Calibrated fairness}}  \cite{saxena2019fairness}} It refers to the balance between providing equal opportunities for all individuals and accommodating for their differences and needs. For example, in social media, a \textit{calibrated fair} algorithm could ensure that all users have equal access to opportunities, such as visibility, while also taking into consideration specific factors, such as language or location, to provide a personalized experience. In healthcare, a \textit{calibrated fair} algorithm could ensure that all patients have access to the same standard of care while accounting for their age and health status to provide the best possible treatment plan. The goal is to strike a balance between treating everyone the same and taking into account individual differences to provide the most equitable outcomes. Fairness metrics, such as the True Positive Rate Difference (TPRD) \cite{mehrabi2021attributing}, False Positive Rate Difference (FPRD) \cite{yao2021refining}, and Equal Opportunity Difference (EOD) \cite{park2021comparison} can be used to assess the level of calibrated fairness. Other commonly used computational methods to achieve calibrated fairness are:

    \begin{enumerate}[i]
        \item {Pre-processing}: transforming the original data set to remove or reduce the effect of sensitive attributes (such as race and gender) on the outcome of a machine learning model \cite{xu2022algorithmic}.
        \item {In-processing}: incorporating fairness constraints into the training process of the model to ensure that the model is calibrated with respect to the sensitive attributes \cite{xu2022algorithmic}.
        \item {Post-processing}: adjusting the output of a model after it has been trained to ensure that it is calibrated with respect to the sensitive attributes \cite{xu2022algorithmic}.
        \item {Adversarial training}: training the machine learning model on adversarial examples, or examples that are specifically designed to challenge the model's ability to make fair predictions \cite{tao2022ruler}.

    \end{enumerate}

    \subsubsection{{\textbf{Statistical fairness}}} It takes into account various factors, such as demographic information, that may be relevant to the notion of fairness in a specific context. Some commonly used statistical definitions of fairness include demographic parity, equal opportunity, and equal treatment \cite{chouldechova2020snapshot}. The \textit{‘demographic parity’} measure can be utilized to minimize data bias by augmenting matrix-factorization objectives with penalty functions \cite{yao2017beyond}, while the \textit{‘equal opportunity’} metric is important to ensure decisions are free from bias \cite{zhang2019fairness}. In the context of social media, individual definitions of fairness might include issues such as unbiased content moderation, fair representation of diverse perspectives and voices, and transparency in the algorithms used to curate and rank content. Commonly used computational metrics are:

    \begin{enumerate}[i]
        \item {Equalized odds}: measures fairness by comparing the true positive rate and false positive rate for different groups \cite{ghassami2018fairness}.
        \item {Theorem of equal treatment}: measures fairness by comparing the treatment of similar individuals belonging to different groups \cite{malawski2020note}.

    \end{enumerate}

    \subsubsection{{\textbf{Intersectional fairness}} \cite{ghosh2021characterizing}} This metric takes into account multiple and intersecting aspects of identity, such as race, gender, and socio-economic status, when making decisions about people. The goal is to ensure that people are not discriminated against, as these intersections can compound and result in greater marginalization and unequal treatment. In the context of social media, an algorithm that takes into account intersectional fairness would ensure that content is not recommended or censored in a biased manner based on a user's race, gender, and socio-economic status, while in the context of healthcare, an algorithm that considers intersectional fairness would ensure that medical treatments and resources are not disproportionately allocated. The best approach to implementing intersectional fairness is through the \textit{worst-case disparity} method. This entails assessing each subgroup individually and comparing the best and worst outcomes to determine the accuracy of the fairness score. The ratio of the maximum and minimum scores is then calculated, and the closer the ratio is to 1, the fairer the outcome is  \cite{ghosh2021characterizing}. Other commonly used methods include:

    \begin{enumerate}[i]
        \item Constraints-based methods: the algorithm is designed to respect certain fairness constraints, such as equal treatment for different groups based on multiple attributes through mathematical optimization \cite{zafar2019fairness}.

        \item Causal inference methods: ensure that the algorithm's outputs are not biased by considering the causal relationships between the inputs and outputs \cite{chakraborti2020contrastive}.

        \item Decision trees and rule-based systems: to ensure that the algorithm's decisions are based on appropriate factors and are not biased \cite{rosenfeld2019explainability}.

    \end{enumerate}

The supervised ranking, unsupervised regression, and reinforcement aspects of fairness evaluation can be done using \textit{pairwise evaluation} \cite{narasimhan2020pairwise}. This involves evaluating the performance of an AI model by comparing its output to a set of predefined pairs of input data.

\begin{table}
\centering
\caption{Fairness evaluation metrics with mathematical formulation. \textit{FP = False Positive, FN = False Negative, TP = True Positive, TN = True Negative, A = binary attribute representing a demographic group}}

\resizebox{0.99\columnwidth}{!}{\begin{tabular}{lll} 
\hline
\textbf{Metric}                 & \textbf{Formula}                                                                                                                                                                                                                                                                                           & \textbf{Description}                                                                                                                                            \\ 
\hline
Equal Opportunity \cite{chouldechova2020snapshot}              & \begin{tabular}[c]{@{}l@{}}$P(\frac{FP}{y=1}) = P(\frac{FP}{A=1, y=1})$\\$- P(\frac{FP}{A=0, y=1})$\\ where y is the true label\end{tabular}                                                                                                                                                               & \begin{tabular}[c]{@{}l@{}}An AI model's positive outcomes are not systematically \\skewed towards or against certain groups of people.\end{tabular}            \\ 
\hline
Equal Odds \cite{chouldechova2020snapshot}                     & \begin{tabular}[c]{@{}l@{}}$P(\frac{y=1}{p\textgreater{}t,y=1}) = P(\frac{y=1}{p\textgreater{}t,y=0})$\\$ = P(\frac{y=0}{p\textless{}=t,y=1}) = P(\frac{y=0}{p\textless{}=t,y=0})$\\ where y is the true label, p \\is the predicted probability \\of positive class, and t is a \\threshold.\end{tabular} & \begin{tabular}[c]{@{}l@{}}The false positive rate and the false negative rate are \\equal across different groups of people.\end{tabular}                      \\ 
\hline
Demographic Parity \cite{chouldechova2020snapshot}             & \begin{tabular}[c]{@{}l@{}}$P(\frac{y=1}{A=1}) = P(\frac{y=1}{A=0})$\\ where y is the predicted label.\end{tabular}                                                                                                                                                                                        & \begin{tabular}[c]{@{}l@{}}The proportion of positive outcomes for different groups \\of people is equal.\end{tabular}                                          \\ 
\hline
Statistical Parity \cite{hertweck2021moral}             & $P(\frac{Y=1 }{ A=a}) = P(Y=1)$ for all a in A                                                                                                                                                                                                                                                             & \begin{tabular}[c]{@{}l@{}}The proportion of favorable outcomes is the same for \\all groups.\end{tabular}                                                      \\ 
\hline
Accuracy \cite{hossin2011hybrid}                       & $\frac{TP + TN}{Total Population}$                                                                                                                                                                                                                                                                         & The proportion of all predictions that are correct.                                                                                                             \\ 
\hline
False Positive Rate (FPR) \cite{yao2021refining}      & $\frac{FP}{FP+TN}$                                                                                                                                                                                                                                                                                         & \begin{tabular}[c]{@{}l@{}}The proportion of negative instances that are incorrectly \\classified as positive.\end{tabular}                                     \\ 
\hline
False Negative Rate (FNR) \cite{nguyen2021leveraging}      & $\frac{FN}{FN+TP}$                                                                                                                                                                                                                                                                                         & \begin{tabular}[c]{@{}l@{}}The proportion of positive instances that are incorrectly\\classified as negative.\end{tabular}                                      \\ 
\hline
True Positive Rate (TPR) \cite{mehrabi2021attributing}       & $\frac{TP}{TP+FN}$                                                                                                                                                                                                                                                                                         & \begin{tabular}[c]{@{}l@{}}The proportion of positive instances that are correctly \\classified as positive. Also known as sensitivity or recall.\end{tabular}  \\ 
\hline
True Negative Rate (TNR) \cite{nguyen2021leveraging}       & $\frac{TN}{TN+FP}$                                                                                                                                                                                                                                                                                         & \begin{tabular}[c]{@{}l@{}}The proportion of negative instances that are correctly \\classified as negative. Also known as specificity.\end{tabular}            \\ 
\hline
Positive Predictive Value (PPV) \cite{xiong2020evaluating} & $\frac{TP}{TP+FP}$                                                                                                                                                                                                                                                                                         & \begin{tabular}[c]{@{}l@{}}The proportion of instances that are predicted as positive \\that are actually positive.\end{tabular}                                \\ 
\hline
Negative Predictive Value (NPV) \cite{xiong2020evaluating} & $\frac{TN}{TN+FN}$                                                                                                                                                                                                                                                                                         & \begin{tabular}[c]{@{}l@{}}The proportion of instances that are predicted as negative \\that are actually negative.\end{tabular}                                \\ 
\hline
False Discovery Rate (FDR) \cite{gu2022electricity}     & $\frac{FP}{FP+TP}$                                                                                                                                                                                                                                                                                         & \begin{tabular}[c]{@{}l@{}}The proportion of instances that are predicted as positive \\that are actually negative.\end{tabular}                                \\ 
\hline
False Omission Rate (FOR) \cite{markoulidakis2021multi}      & $\frac{FN}{FN+TN}$                                                                                                                                                                                                                                                                                         & \begin{tabular}[c]{@{}l@{}}The proportion of instances that are predicted as negative \\that are actually positive.\end{tabular}                                \\ 
\hline
Positive Likelihood Ratio (LR+) \cite{vergeer2021measuring} & $\frac{TP}{FP}$                                                                                                                                                                                                                                                                                            & \begin{tabular}[c]{@{}l@{}}Indicates how much more likely a positive result is to occur \\when the condition is present than when it is absent.\end{tabular}    \\ 
\hline
Negative Likelihood Ratio (LR-) \cite{vergeer2021measuring} & $\frac{FN}{TN}$                                                                                                                                                                                                                                                                                            & \begin{tabular}[c]{@{}l@{}}Indicates how much more likely a negative result is to occur \\when the condition is absent than when it is present.\end{tabular}    \\
\hline
\end{tabular}}
\end{table}

\section{An Overview of Accountability}\label{accountability}

\subsection{Definitions, Computational Methods, and Approaches to Accountability}

It refers to the notion that individuals or organizations using AI should be responsible and answerable for the consequences of their systems. This includes the responsibility to ensure that the system operates in an ethical manner, with the goal of providing equitable and accurate outcomes. There are several definitions of accountability in AI, including:

\subsubsection{{\textbf{Legal accountability} \cite{zaki2021supporting}}} It refers to the legal obligations of entities involved in the design, development, deployment, and use of AI systems for healthcare purposes in social media. This includes the responsibility for ensuring that the AI systems are developed and used in accordance with applicable laws and regulations, as well as the responsibility for any negative consequences or impacts that may result from their use. Legal accountability may also extend to issues such as data protection and privacy, and the responsibility for ensuring that AI systems are not used for discriminatory or unethical purposes. The commonly used conceptual computational methods are:
    \begin{enumerate}[i]
        \item Transparency: Ensuring that AI systems are transparent and that their decision-making processes can be explained and understood \cite{blacklaws2018algorithms}.

        \item Documentation: Keeping records of systems' design, development, and testing processes, as well as the data used to train them \cite{dubberley2020digital}.

        \item Auditing: Conducting independent assessments of AI performance and accuracy to verify compliance with legal requirements \cite{parker2007meta}.

        \item Regulation: Implementing legal frameworks that establish standards and requirements for the development, deployment, and use of AI systems \cite{parker2007meta}.

        \item Adjudication: Establishing procedures for resolving disputes and grievances related to the use of AI systems \cite{king2013instrumental}.

    \end{enumerate}

    \subsubsection{ \textbf{Ethical accountability} \cite{mittelstadt2019principles}} It involves ensuring that the decisions made by AI systems are transparent, justifiable, and in line with the values of society. This includes issues such as data privacy, informed consent, and ensuring that AI systems do not perpetuate existing biases and discrimination. The ethical considerations around the use of AI in healthcare include topics such as the protection of patient privacy, the use of sensitive health data, and the potential for AI systems to reinforce existing health disparities \cite{kass2018ethics}. Stakeholders involved in the development and deployment of AI in healthcare have a responsibility to ensure that ethical principles are integrated into the design and implementation of these systems, and that the outcomes of their use are regularly monitored and evaluated for any ethical concerns. Some common methods include:
    \begin{enumerate}[i]
        \item Ethical Impact Assessment: involves identifying the ethical risks and benefits of the system, and determining the trade-offs between them \cite{wright2011framework}.
    
        \item Value Alignment: invloves incorporating ethical principles and values into the design and development of the system, and ensuring that its behavior is consistent with these values. \cite{arnold2017value}.

        \item Transparency and Explanation: achieved by providing clear and concise explanations of how the system works, and by making its data and algorithms open and accessible \cite{iyer2018transparency}.

        \item Stakeholder Engagement: involves engaging stakeholders, including users, developers, and experts, in the development and evaluation of AI systems \cite{unerman2010stakeholder}.

    \end{enumerate}

    \subsubsection{\textbf{Technical accountability} \cite{wachter2017transparent}} It refers to the responsibility of the developers and designers of AI systems to ensure that their technology meets certain standards of functionality, security, and privacy. This includes having appropriate systems in place to monitor and manage the AI algorithms, as well as addressing any technical issues that arise. In the context of social media and healthcare, technical accountability also involves considering how AI technologies can be used to support ethical decision-making, such as ensuring that user privacy is protected and that decisions are made in a fair and transparent manner \cite{ozga2020politics}. Some of the commonly used methods:

    \begin{enumerate}[i]
        \item Logging: logging all inputs, outputs, and decisions can be used to track the system's performance and identify potential issues \cite{ko2011system}.

        \item Auditing: to assess their performance, identify potential biases, and ensure that they are aligned with ethical and legal standards \cite{raji2020closing}.

        \item Transparency, Model interpretability, and Explainability: can be designed to provide users with clear explanations of their decision-making processes, which can help to increase trust in the system and reduce the risk of ethical and legal violations \cite{wachter2017transparent}.

    \end{enumerate}

    \subsubsection{\textbf{Societal accountability} \cite{vesnic2020societal}} It refers to the responsibility of the stakeholders to ensure that the use of AI systems aligns with the values and interests of society as a whole. This includes issues such as privacy, transparency, and fairness, as well as broader social, cultural, and economic factors that can be affected by AI systems. To ensure societal accountability, it may be necessary for stakeholders to engage in public consultation, to develop regulations and standards that ensure that AI systems are used ethically and transparently, and to promote transparency and public understanding of how AI systems work and what they are being used for. Ultimately, it means that the development and use of AI systems is guided by the principles of responsible innovation, and that the interests of society are taken into account in all stages of their lifecycle. Other methods are:
    \begin{enumerate}[i]
        \item Regulation and standardization: development of regulations and standards for the design and use of AI systems. This can help ensure that AI systems are accountable to society and that they operate in a way that protects the rights and interests of all stakeholders \cite{kerikmae2021legal}.

        \item Public-private partnerships: collaboration between government agencies, private companies, and civil society organizations to ensure that AI systems are developed and used in a way that is accountable to society \cite{reich2018core}.

    \end{enumerate}

Accountability can be ensured by implementing transparency and fairness into the algorithms, designing systems with privacy in mind, and conducting regular audits and evaluations to assess the performance of the AI system. Researchers have proposed a method for holding companies accountable for their actions related to AI \cite{wieringa2020account}. They argue that it is crucial to first identify the specific decision-makers within the company who are responsible for the error in question. This is essential for ensuring fair judgment. The person or group responsible for determining accountability should be well-versed in the various legal, political, administrative, professional, and social perspectives related to the topic of the error to ensure that the judgement is fair and unbiased. Finally, the consequences for the decision-makers should be tailored to the specific areas of their responsibility, and the level of responsibility of each individual decision-maker within the company's hierarchy should be considered when determining them. Algorithms such as decision trees and regression models are more interpretable than others \cite{slack2019assessing}. With the widespread adoption of deep learning methods in decision models, explainability in AI (XAI) also presents a way to interpret the model by humans.

\begin{table}[]
\caption{Accountability evaluation metrics with mathematical formulation. Accuracy, False Positive Rate, and False Negative Rate metrics are also suitable for evaluating accountability in AI systems, as discussed in Table \ref{table1}. \textit{FP = False Positive, FN = False Negative, TP = True Positive, TN = True Negative}}

\resizebox{0.99\columnwidth}{!}{\begin{tabular}{lll}
\hline
\textbf{Metric}                     & \textbf{Formula}                                                                                                          & \textbf{Description}                                                                                            \\ \hline
Fairness \cite{lagioia2022algorithmic}                           & \begin{tabular}[c]{@{}l@{}}${\fontsize{50}{60}\frac{\#TP\_ for\_ group A }{ \#actual\_ positives\_ for\_  A}}$ \\ $- \frac{\#TP\_ for\_ group B}{\#actual\_ positive\_ for\_  B}$ \end{tabular} & \begin{tabular}[c]{@{}l@{}}Measures whether the AI system treats different groups \\fairly \end{tabular}                                                   \\ \hline
Explainability \cite{kaur2021trustworthy}                     & $\frac{\#explanations\_ provided}{\# decisions\_ made}$                                                                      & \begin{tabular}[c]{@{}l@{}}Measures how well the AI system can explain its \\decision-making process \end{tabular}                                        \\ \hline
Consistency \cite{bucher2016improving}                        & $1 - \frac{\# changes\_ to\_ output}{\# decisions\_ made}$                                                                     & \begin{tabular}[c]{@{}l@{}}Measures how consistent the AI system's outputs \\are over time \end{tabular}                                                  \\ \hline
Robustness \cite{weng2018evaluating}                         & $\frac{\#correct outputs}{\# decisions made}$                    & \begin{tabular}[c]{@{}l@{}}Measures how well the AI system performs under \\unexpected conditions \end{tabular}                                           \\ \hline

Precision \cite{kynkaanniemi2019improved}                          & $\frac{TP}{TP+FP}$                                                                                                       & \begin{tabular}[c]{@{}l@{}}The proportion of true positive predictions \\among all positive predictions.    \end{tabular}                                 \\ \hline
Recall (Sensitivity) \cite{kynkaanniemi2019improved}              & $\frac{TP}{TP+FN}$                                                                                                       & \begin{tabular}[c]{@{}l@{}}The proportion of true positive predictions \\among all actual positive instances.  \end{tabular}                              \\ \hline
Specificity \cite{jadon2020survey}                        & $\frac{TN}{TN+FP}$                                                                                                       & \begin{tabular}[c]{@{}l@{}}The proportion of true negative predictions \\among all actual negative instances.\end{tabular}                                \\ \hline
F1 Score \cite{grandini2020metrics}                           & $2 * \frac{Precision * Recall}{ Precision + Recall}$                                                                        & \begin{tabular}[c]{@{}l@{}}The harmonic mean of precision and \\recall. \end{tabular}                                                                     \\ \hline
                            
Confusion Matrix \cite{markoulidakis2021multi}                   & ${[}{[}TP,FP{]}, {[}FN,TN{]}{]}$                                                                                            & A table used to evaluate the performance of a classifier.                                                       \\ \hline
Pandora \cite{nushi2018towards}                   & \begin{tabular}[c]{@{}l@{}}Five-fold cross validation \\for cluster prediction accuracy \end{tabular}                                                               & \begin{tabular}[c]{@{}l@{}}It is a hybrid of human and system-generated observations \\to explain system failure for analysis and debugging.\end{tabular} \\ \hline
\end{tabular} }
\end{table}

\section{An Overview of Transparency}\label{transparency}

\subsection{Definitions, Computational Methods, and Approaches to Transparency}

Transparency in AI refers to the degree to which the internal workings of an AI system can be understood by humans \cite{adadi2018peeking}. It involves providing explanations for how the system makes decisions, understanding the data that was used to train the system, and ensuring that the system is not biased or discriminatory. The issues of transparency and privacy are often at cross-heads. For example, while analyzing mental health data on social media platforms, the challenge is not with the identifying attributes of individual users (as the data is often aggregated), but with how that data is utilized \cite{conway2016social}. There are several different definitions of transparency depending on the specific context and use case:

\subsubsection{\textbf{Algorithmic transparency} \cite{diakopoulos2017algorithmic}} It refers to the ability to understand how an AI algorithm or model arrives at its outputs or decisions. In the context of social media for healthcare, transparency can be defined as the ability to clearly understand the processes and methods used to create, disseminate, and evaluate social media interventions for healthcare purposes \cite{stellefson2020evolving}. This includes being able to understand the data sources used to inform the interventions, the algorithms or models used to analyze the data and create the interventions, and the criteria used to evaluate the effectiveness of the interventions. Transparency is important because it allows for the identification and mitigation of potential biases or errors in the interventions, and helps to build trust with stakeholders, including patients, healthcare providers, and regulators. There are several computational methods that can be used to increase algorithmic transparency, such as:

    \begin{enumerate}[i]
        \item Feature importance analysis: involves identifying the features or variables that have the most significant impact on the model's output. By doing so, it helps to understand the model's decision-making process \cite{valko2010feature}.
        \item Model interpretability: involves designing models in such a way that their output can be easily understood and interpreted by humans. For example, decision trees are considered interpretable because their output can be visualized as a series of decision nodes \cite{lipton2018mythos}.
        \item Explanation generation: involves generating explanations for the model's output. These explanations can be in the form of natural language or visualizations, and they help to provide insight into the model's decision-making process \cite{stepin2021survey}.

    \end{enumerate}

    \subsubsection{\textbf{ Data transparency}} It refers to the ability to understand how data is collected, stored, and used in the development of an AI system \cite{bertino2019redefining}. In the context of healthcare, data transparency refers to the extent to which healthcare organizations and providers are open and clear about the collection, storage, and use of patient data in the design and implementation of their social media campaigns \cite{he2019practical}. This includes providing patients with clear information about what data is being collected, how it will be used, who will have access to it, and how it will be protected. By being transparent about data collection and use, healthcare organizations can build trust with patients and promote more active engagement in social media-based health interventions. This can ultimately lead to better health outcomes for patients, as they are more likely to participate in interventions that they feel comfortable with and have confidence in. Computational methods for data transparency can include:
    \begin{enumerate}[i]
        \item Data visualization: involves creating graphical representations of data in order to make it easier for users to understand and interpret \cite{li2020overview}. 
        \item Data profiling: involves analyzing data to understand its structure, quality, and content, which can help identify issues such as missing values or inconsistencies \cite{azeroual2018analyzing}. 
        \item Data lineage analysis: involves tracing the movement of data through various systems and processes to ensure its accuracy and reliability \cite{bose2005lineage}.

    \end{enumerate}
    
    \subsubsection{\textbf{Process transparency}} It refers to the ability to understand the steps taken to develop and deploy an AI system, including the testing and validation processes used \cite{leslie2019understanding}. In the context of social media and healthcare, it refers to the transparency of the decision-making process that determines which health-related information is prioritized, displayed, and disseminated on social media platforms \cite{world2016medicines}. This can include transparency around the algorithms and other computational methods used to curate and display health-related content, as well as the policies and guidelines used to moderate user-generated content related to health. By increasing process transparency, users can have more confidence in the information and interventions being presented to them, and researchers can have greater trust in the data they are analyzing. There are several computational methods that can be used to increase process transparency in AI systems:

    \begin{enumerate}[i]
        \item Data provenance tracking: involves tracking the origin, processing history, and movement of data throughout the AI system. This helps to ensure that the data used in the system is reliable and can be traced back to its source \cite{burgess2022visualizing}.
        \item Model interpretability: involves developing algorithms and tools that can help explain how an AI system makes decisions. Techniques such as feature importance analysis \cite{valko2010feature}, decision trees \cite{fan2011hybrid}, and partial dependence plots \cite{zhao2020modelling} can help to uncover how the model arrives at its predictions.
        \item Explanation generation:  involves generating natural language or visual explanations for the decisions made by an AI system. Techniques such as saliency maps, LIME (Local Interpretable Model-agnostic Explanations) \cite{zafar2021deterministic}, and SHAP (SHapley Additive exPlanations) \cite{lundberg2017unified} can help to generate these explanations.
        \item Auditability and monitoring: involves building auditing and monitoring capabilities into the AI system. This can include monitoring the system's performance, detecting bias or other ethical issues, and identifying when the system is not performing as intended \cite{shneiderman2020bridging}.
        \item Open-source development: involves developing AI systems in an open and transparent manner, where the code, data, and models are publicly available. This allows for greater scrutiny and accountability of the system by external stakeholders, such as regulators or the general public \cite{brundage2020toward}.

    \end{enumerate}
    
    \subsubsection{\textbf{ Explainability}} It refers to the ability to provide a clear and understandable explanation of how an AI system arrived at a particular decision or recommendation \cite{janssen2022will}. In the context of social media intervention for healthcare, explainability can involve understanding how an AI system is processing social media data, how it is identifying relevant information, and how it is making recommendations or decisions based on that data \cite{paredes2022heic}. It can also involve understanding the factors that influenced the system's decision-making, such as the data used to train the model or the specific features that were weighted more heavily in the decision process. To achieve explainability in AI systems for social media intervention in healthcare, various methods can be used, including techniques for feature selection, model interpretability, and visualizations. These methods can help healthcare professionals to better understand the underlying mechanisms of an AI system and the factors that contribute to its decision-making process. Some common methods include:

    \begin{enumerate}[i]
        \item Decision trees: Decision trees are graphical representations of the decision-making process of a model. They can be used to explain how the model is making decisions and which factors are most influential \cite{fan2011hybrid}.
        \item LIME (Local Interpretable Model-Agnostic Explanations): LIME is a method for explaining the predictions of any machine learning model. It works by generating a simpler, interpretable model that approximates the behavior of the original model \cite{zafar2021deterministic}.
        \item SHAP (SHapley Additive exPlanations): SHAP is a method for explaining the output of any machine learning model. It works by computing the contribution of each input feature to the final prediction \cite{lundberg2017unified}.
        \item Counterfactual explanations: Counterfactual explanations involve identifying the minimal set of changes to the input features that would result in a different output from the model. They can be used to explain why a certain prediction was made and what could have been done differently to change the outcome \cite{sokol2019counterfactual}.

    \end{enumerate}

    \subsubsection{\textbf{Interpretability}} It refers to the ability to understand the meaning and implications of the decisions made by an AI system, including how they impact different groups of people \cite{carvalho2019machine}. Interpretability also refers to the ability of an AI system to provide a clear and understandable explanation for its decisions or recommendations to healthcare professionals, patients, and other stakeholders \cite{amann2020explainability}. This is particularly important in healthcare, where the consequences of AI decisions can be critical to patient outcomes. An interpretable system enables stakeholders to understand how the AI arrived at its recommendation and can help build trust in the system. Interpretability techniques in AI involve designing models with clear and understandable features, such as decision trees or rule-based systems \cite{arrieta2020explainable}. These methods can help identify the factors that influenced the AI's decision, making it easier to understand and explain the outcome. Other techniques include generating visualizations, such as heatmaps or saliency maps, which highlight the areas of an input that had the most significant impact on the model's output. By providing clear explanations of the model's decision-making process, these techniques can help stakeholders better understand and trust the AI system. Some of the computational methods are:

    \begin{enumerate}[i]
        \item Partial Dependence Plot (PDP): PDP shows the relationship between the target variable and one or two input variables while controlling for the effects of other input variables. This shows how the model is making predictions and how the input variables are affecting the output \cite{zhao2020modelling}.
        \item Local Interpretable Model-Agnostic Explanations (LIME): LIME is a post-hoc method that explains the output of any classifier by approximating it with an interpretable model locally. This shows how the model is making decisions for a specific instance \cite{zafar2021deterministic}.
        \item Model Distillation: It is the process of training a simpler model that approximates the decision boundaries of a more complex model. This can help in creating a simpler and more interpretable model that still maintains the performance of the original model \cite{murdoch2019definitions}.
    \end{enumerate}

Overall, transparency in AI is important for ensuring accountability, fairness, and ethical use of AI systems. It helps build trust with users and stakeholders, and can also help identify and address biases or errors in the system.

\begin{table}[]
\caption{Transparency evaluation metrics with mathematical formulation}

\resizebox{0.99\columnwidth}{!}{\begin{tabular}{lll}
\hline
\textbf{Metric} & \textbf{Formula}                                                                     & \textbf{Description}                                          \\ \hline
Completeness \cite{weiskopf2013defining}   & $\frac{\#available\_ data\_ points}{\# total\_ data\_ points}$                                   & \begin{tabular}[c]{@{}l@{}}The extent to which all relevant information \\is available\end{tabular}     \\ \hline
Timeliness \cite{crawley2021using}     & $\frac{\# data\_ points\_ available\_ within\_ timeframe}{\# required\_ data\_ points}$ & \begin{tabular}[c]{@{}l@{}}The extent to which data is available in a \\timely manner\end{tabular}      \\ \hline
Relevance \cite{zhai2015beyond}      & $\frac{\# relevant\_ data\_ points}{\# data\_ points}$                                          & \begin{tabular}[c]{@{}l@{}}The extent to which data is applicable to \\the problem at hand\end{tabular} \\ \hline
Accessibility \\ \cite{weiss2018global}  & $\frac{\# data\_ points\_ that\_ can\_ be\_ obtained}{\# data\_ points}$                     & \begin{tabular}[c]{@{}l@{}}The extent to which data is easy to obtain \\and use\end{tabular}            \\ \hline
Data Provenance \cite{burgess2022visualizing} & $\frac{\% data\_ with\_ known\_ source}{chain\_ of\_ custody}$                                        & \begin{tabular}[c]{@{}l@{}}Involves tracking the origin, processing history, and \\movement of data throughout the AI system. \end{tabular}                                                              \\ \hline
\end{tabular}}
\end{table}

\section{An Overview of Ethics}\label{ethics}

\subsection{Definitions, Computational Methods, and Approaches to Ethics}
 In AI, ethics refers to the study and practice of developing and implementing AI technologies in a manner that is fair, transparent, and beneficial to all stakeholders \cite{leikas2019ethical}. The goal of ethical AI is to ensure that AI systems and their decisions are aligned with human values, respect fundamental human rights, and do not result in harm or discrimination against individuals or groups. This includes considerations of privacy, data protection, bias, accountability, and explainability \cite{latonero2018governing}. In the context of social media, digital surveillance of public health data from social media platforms should be guided by the principles of 1) beneficence: surveillance must lead to improvement in public health outcomes; 2) non-maleficence: use of data should not erode public trust; 3) autonomy: informed consent of users or anonymizing of identifying details; 4) equity: equal opportunities to individuals for public health interventions, and 5) efficiency: building legal mandates to ensure continuous access to web platforms and decision-making algorithms \cite{aiello2020social}. AI-mediated healthcare treatments must account for affordability and equality among the masses, and while nascent, health tech in the field of \textit{patient-centric} models is no longer science fiction, wherein scientifically tailored medicines are prescribed to the patients \cite{gomez2020artificial}. There are many different definitions of ethics, depending on the context in which the term is used: 

\subsubsection{\textbf{Philosophical ethics}}  Refers to the concept of ensuring that AI systems are designed and used in ways that respect human autonomy, dignity, and privacy \cite{kazim2021high}. In the context of social media intervention for healthcare, philosophical ethics in AI refers to the study and application of ethical principles and values to the development and use of AI-powered tools and technologies for healthcare interventions via social media \cite{nebeker2022ai}. It involves examining the potential benefits and risks of using AI to collect, analyze, and interpret health-related data from social media platforms, as well as ensuring that the use of such technologies aligns with ethical principles such as respect for privacy, autonomy, beneficence, and non-maleficence. It also involves considering the potential biases that may arise in the development and use of these technologies, and taking steps to mitigate these biases to ensure that the use of AI in social media intervention for healthcare is fair, just, and equitable for all individuals involved. Ultimately, the aim is to promote the development and use of AI technologies that improve health outcomes, while minimizing the potential risks and harms that may arise from their use. Some examples of computational methods for philosophical ethics include:

    \begin{enumerate}[i]
        \item Simulation and modeling: These are techniques that allow ethical dilemmas to be simulated and modeled, providing insights into the likely outcomes of different ethical decisions \cite{crockett2013models}.
        \item Game theory: This is a mathematical framework that can be used to model and analyze decision-making in social situations, including ethical dilemmas \cite{sanfey2007social}.
        \item Data analytics: This involves the use of statistical methods and machine learning algorithms to analyze data and identify patterns or insights related to ethical questions or dilemmas \cite{someh2019ethical}.

    \end{enumerate}

    \subsubsection{\textbf{Moral ethics}} Refers to the ethical considerations that need to be taken into account while using social media for healthcare interventions \cite{guttman2017ethical}. This includes ensuring that the privacy and confidentiality of patient data are maintained, that the patient's autonomy and consent are respected, and that the use of social media platforms does not result in any harm to the patient \cite{bhatia2019ethical}. It also involves ensuring that the interventions are based on evidence-based practices and that the potential benefits of the intervention outweigh the potential risks. Finally, it involves being transparent about the use of social media for healthcare interventions and communicating the risks and benefits to all stakeholders involved \cite{sorensen2018health}. Some of the computational methods are:

    \begin{enumerate}[i]
        \item Data visualization tools: These tools can be used to present complex ethical data in a clear and accessible way, making it easier for healthcare professionals and other stakeholders to understand and make informed decisions \cite{davis2012ethics}.
        \item Sentiment analysis: Language and sentiment of social media posts related to healthcare interventions can help identify any ethical issues or concerns that may arise, such as biases or stigmatization of certain patient groups \cite{livingston2012effectiveness}.
        \item Crowdsourcing platforms: developed to gather feedback from a diverse group of individuals on the ethical implications of the AI system and its recommendations. This can help ensure that the system takes into account a range of perspectives and values, and can identify potential ethical concerns that may have been overlooked by the development team \cite{jakesch2022different}.

    \end{enumerate}

    \subsubsection{\textbf{Professional ethics}} In the context of healthcare and social media intervention, professional ethics involves a set of guidelines and principles that guide the behavior of HCPs who are using social media as part of their practice \cite{ventola2014social}. This might include guidelines around patient privacy, confidentiality, informed consent, and the appropriate use of social media platforms (i.e., avoiding conflicts of interest or biased behavior) for sharing health information. Computational methods for enforcing professional ethics might include:

    \begin{enumerate}[i]
        \item Automated systems: for monitoring healthcare professionals' behavior on social media platforms \cite{swan2009emerging}.
        \item Algorithms to detect and flag any instances of inappropriate behavior or violations of professional ethical standards \cite{drabiak2020should}.

    \end{enumerate}

    \subsubsection{\textbf{Social ethics}} Refers to the moral principles and values that would involve considerations of how the use of social media affects the privacy, autonomy, and well-being of patients and other stakeholders, as well as issues related to fairness and equity \cite{wright2011framework}. For example, social ethics would require that healthcare providers and organizations respect the privacy of patients and protect their personal information when using social media platforms \cite{gagnon2015professionalism}. Social ethics would also require that healthcare providers and organizations take steps to ensure that the use of social media in healthcare does not create or reinforce existing health disparities, such as by providing access to care or health information only to certain groups of people who have access to social media. Overall, it provides a framework for evaluating the social and moral implications of using social media for healthcare interventions and for ensuring that these interventions are conducted ethically and responsibly. Several methods can contribute to the promotion of social ethics in AI:

    \begin{enumerate}[i]
        \item Fairness-aware machine learning algorithms: These aim to mitigate unfairness in the training data and algorithmic decision-making process \cite{pastaltzidis2022data}.
        \item Privacy-preserving data analysis: These aim to protect sensitive data from unauthorized access, while still allowing for meaningful analysis \cite{keshk2022privacy}.
        \item Human-in-the-loop approaches: These incorporate human oversight and decision-making into the AI system, to ensure that the system is aligned with social values and ethical principles \cite{enarsson2022approaching}.
        \item Explainable AI: This is a computational method that aims to make AI systems more transparent and understandable to users, so that they can make informed decisions about the ethical implications of the system's output \cite{adadi2018peeking}.
        \item Value-sensitive design: This seeks to identify and incorporate social values and ethical principles into the design and development of AI systems, in order to promote their alignment with social ethics \cite{umbrello2021mapping}.

    \end{enumerate}

    \subsubsection{\textbf{Legal ethics}} Refers to the ethical considerations related to complying with the laws, regulations, and policies surrounding healthcare data privacy and security \cite{stanberry2006legal}. This includes maintaining the confidentiality of patient data, adhering to informed consent and data-sharing agreements, and complying with relevant legal and ethical standards \cite{kalkman2019responsible}. It also involves ensuring that the AI models used in social media intervention for healthcare are developed and used in compliance with relevant regulations and standards. Legal tools for ensuring ethics include:
    \begin{enumerate}[i]
        \item HIPAA (Health Insurance Portability and Accountability Act): implementing privacy regulations for healthcare data \cite{hansen1997hipaa}.
        \item GDPR (General Data Protection Regulation): complying with data protection laws and adhering to other relevant legal and regulatory frameworks that govern the use of AI in healthcare and social media interventions \cite{sharma2019data}.
        \item Ethical Research Board (ERB:)The idea of Ethics by Design suggests incorporating the services of an ERB while developing any product in an organization \cite{leidner2017ethical}.

    \end{enumerate}

\begin{table}
\caption{Ethics evaluation metrics with mathematical formulation. \textit{FP = False Positive, FN = False Negative, TP = True Positive, TN = True Negative}}

\resizebox{0.99\columnwidth}{!}{\begin{tabular}{lll}
\hline
\multicolumn{1}{c}{\textbf{Metric}} & \multicolumn{1}{c}{\textbf{Formula}}                                                            & \multicolumn{1}{c}{\textbf{Description}}                                                           \\ \hline
Bias \\ \cite{dixon2018measuring}                               & $(\frac{\#FP\_for\_ Group A}{\# TN \_for\_ Group A})/(\frac{\#FP\_for\_ Group B}{\#TN\_ for\_ Group B})$                      & \begin{tabular}[c]{@{}l@{}}Measures the extent to which an AI system exhibits \\bias towards a particular group or demographic.\end{tabular} \\ \hline
Discrimination \\ \cite{pencina2012novel}                     & $\frac{P(positive\_ outcome\_ for\_ Group A)}{ P(positive\_ outcome\_ for\_ Group B)}$                               & \begin{tabular}[c]{@{}l@{}}Measures whether an AI system is treating different \\groups of people unfairly. \end{tabular}                    \\ \hline
Privacy \cite{mendes2017privacy}                            & $\frac{\# privacy\_ violations}{\# individuals\_ whose\_ data\_ was\_ processed}$                              & \begin{tabular}[c]{@{}l@{}}Measures the extent to which an AI system is \\protecting the privacy of individuals. \end{tabular}                \\ \hline
Accountability \cite{link2012public}                     & $\frac{\# system\_ was\_ found\_ to\_ be\_ at\_ fault}{ \#interactions\_ with\_ system}$                               & \begin{tabular}[c]{@{}l@{}}Measures whether an AI system can be held \\accountable for its actions.\end{tabular}                             \\ \hline
Transparency \cite{datta2016algorithmic}                       & $\frac{\# of\_ decisions\_ that\_ can\_ be\_ explained}{Total \# of\_ decisions\_ made\_ by\_ AI\_ system}$ & \begin{tabular}[c]{@{}l@{}}Ensuring that the decision-making process of \\an AI system is clear and understandable to users.\end{tabular}    \\ \hline
\end{tabular}}
\end{table}

\section{FATE in Data Sets}\label{datasets}

Research has also been undertaken to improve transparency and accountability in the creation and use of datasets \cite{hutchinson2021towards}. Here, the authors propose that thorough documentation should be kept for every step of the process, from the initial design to the final product. This would allow for clear identification of those who are responsible for any errors that may occur. Additionally, each stage of the development process should have a designated leader who takes ownership of their section of the program. Providing explanations for the purpose and function of each section can help to increase understanding and transparency. Furthermore, regular maintenance and documentation of updates should be conducted to not only minimize future mistakes but also boost morale among developers by highlighting progress made. The Adult Income dataset \cite{dua2017uci}, the German Credit dataset \cite{asuncion2007uci}, and the UCI Credit Card dataset \cite{asuncion2007uci} are commonly used while evaluating FATE models.

The performance of AI systems with respect to FATE principles can be evaluated using metrics to ensure that AI systems are making fair and unbiased decisions.

\subsection{\textbf{FATE toolkits}} In recent years, researchers have been motivated to develop AI tools which can detect the level of bias present in a decision. Aequitas \cite{saleiro2018aequitas} produces reports that can facilitate equitable decision-making for policymakers and ML researchers, while the AI Fairness 360 \cite{bellamy2019ai} and Fairlearn \cite{bird2020fairlearn} toolkit provides performance benchmarking for fairness algorithms. These libraries can therefore be used for assessing and mitigating bias in AI models, including methods for data pre-processing, model training, and post-processing.

It's important to note that FATE evaluations are not only a one-time assessment, but a continuous process, where metrics can be used to track the performance of the AI over time. Using these metrics, organizations can ensure that their AI systems are fair, accountable, transparent and ethical, and that they are making decisions that are in the best interest of all individuals.

\section{Discussion}\label{discussion}

Medical corporations use social media to advertise their services, reach out to people and build a sense of community \cite{grajales2014social}. Social media platforms also provide a means for medical professionals to interact with patients and gather feedback, enabling them to improve their services. Social media can also be used to improve health through peer-to-peer encouragement, raise awareness on diseases, and for doctors to reach their patients through online consultations \cite{chen2021social}. To combat misinformation, more fact-checking is needed, and more health institutions need to reach out to patients to ensure they are getting accurate information. The use of social media for health professionals should be carefully monitored to ensure patient confidentiality is maintained.

This study helps us identify the following principal findings:

\begin{itemize}

\item \textbf{RQ1}: What are the existing solutions to FATE (Fairness, Accountability, Transparency, and Ethics) when discussing healthcare on Social Media Platforms (SMPs)?

The existing solutions to FATE when discussing healthcare on SMPs are:
\begin{enumerate}[i]
    \item Healthcare fairness addressed through calibrated, statistical, and intersectional fairness. Calibrated fairness balances equal opportunities with personalized differences like language or location. Statistical fairness considers demographic information to avoid bias. Intersectional fairness considers multiple aspects of identity.

    \item Accountability in healthcare on SMPs involves legal compliance, ethical principles in system design, technical functionality and privacy, and societal regulation and standardization. This involves protecting data privacy, avoiding discriminatory or unethical use of AI systems, conducting ethical impact assessments, promoting transparency, engaging stakeholders, conducting audits and evaluations, and holding decision-makers accountable.

    \item Transparency in AI refers to understanding how an AI system works, including its algorithms, data sources, and decision-making processes. In social media for healthcare, transparency involves understanding how interventions are created, disseminated, and evaluated. Transparency is important for identifying and mitigating biases or errors, building stakeholder trust, and promoting engagement in social media-based health interventions.

    \item Ethics in healthcare on SMPs involves developing fair, transparent, and beneficial AI technologies. This includes addressing privacy, data protection, bias, accountability, and explainability. Professional ethics and social ethics, such as patient privacy and autonomy, are also important. The goal is to promote ethical use of AI in healthcare on SMPs while minimizing risks and harms.
\end{enumerate}

\item \textbf{RQ2}: How do the different solutions identified in response to RQ1 compare to each other in terms of computational methods, approaches, and evaluation metrics?

The different solutions identified in response to RQ1 for healthcare fairness on SMPs can be compared in terms of computational methods, approaches, and evaluation metrics. The solutions include: Calibrated, statistical, and intersectional fairness achieved through various computational methods such as data pre-processing, adversarial training, and decision trees/rules. Evaluation metrics include Equal Opportunity and Equal Odds. The solution for accountability involves regulatory measures and public-private partnerships to ensure transparency, fairness, privacy, and accountability. Transparency in AI can be achieved through algorithmic transparency, data transparency, and process transparency. Algorithmic transparency can be increased through methods like feature importance analysis, model interpretability, and explanation generation. Data transparency can be improved through data visualization, profiling, and lineage analysis. Process transparency can be enhanced through data provenance tracking, interpretability, explanation generation, auditability, monitoring, and open-source development. Ethics in healthcare on SMPs can be promoted through simulation, modeling, data analytics, sentiment analysis, crowdsourcing, and automated systems, while also considering professional ethics and social ethics.

\item \textbf{RQ3}: What is the strength of evidence supporting the different solutions?

The strength of evidence supporting different solutions for healthcare fairness on Social Media Platforms (SMPs) varies depending on factors such as research quality, methodology, and statistical significance. Calibrated fairness, statistical fairness, and intersectional fairness have established concepts with significant research support. Computational methods like data pre-processing, adversarial training, and decision trees/rules are commonly used, but evidence of their effectiveness may vary. Evaluation metrics such as Equal Opportunity and Equal Odds are commonly used but rely on established statistical measures. Ethics in healthcare on SMPs, including privacy protection and bias mitigation, are guided by established principles, but evidence supporting specific solutions may vary. Solutions like simulation, modeling, data analytics, and crowdsourcing are widely used, but their evidence may vary depending on context. Consulting reputable sources for up-to-date research findings is important due to the dynamic nature of the field.
\end{itemize}

\section{Future Research Directions}\label{future}

This study offers a comprehensive overview of the challenges and progress related to FATE in AI. Despite advancements, challenges remain for AI systems in healthcare, including ethical considerations for patient decision-making, accuracy, and understanding of decision-making processes. Obtaining trustworthy data sets and informed user consent, especially for large language models like ChatGPT, which have the potential of being used in clinical settings, are challenging. Overconfidence in AI systems can also lead to skepticism from clinicians. Additionally, the lack of mathematical formulation for many FATE computational methods and approaches creates a gap between computational and evaluation metrics.

\section{Conclusion}\label{conclusion}

The purpose of this review was to provide a comprehensive analysis of FATE solutions in AI for social media and healthcare, and to highlight recent trends and research gaps in the field. By examining the definitions, computational methods, approaches, and data sets used in the literature, we identified both the progress made and the challenges that remain in achieving FATE in AI. Through our evaluation of the papers, we also highlighted the need for researchers to use appropriate evaluation metrics and data sources when analyzing their approaches. While some progress has been made, there is still much work to be done in order to address the remaining challenges. We hope that this review will serve as a useful resource for researchers and stakeholders, and that it will encourage further research in this important area. Ultimately, our goal is to support the development of FATE-ready AI systems that can be deployed ethically and responsibly in social media and healthcare.

\printcredits

%% Loading bibliography style file
% \bibliographystyle{model1-num-names}
\bibliographystyle{cas-model2-names}

% Loading bibliography database
\bibliography{cas-refs}

\begin{thebibliography}{139}
\expandafter\ifx\csname natexlab\endcsname\relax\def\natexlab#1{#1}\fi
\providecommand{\url}[1]{\texttt{#1}}
\providecommand{\href}[2]{#2}
\providecommand{\path}[1]{#1}
\providecommand{\DOIprefix}{doi:}
\providecommand{\ArXivprefix}{arXiv:}
\providecommand{\URLprefix}{URL: }
\providecommand{\Pubmedprefix}{pmid:}
\providecommand{\doi}[1]{\href{http://dx.doi.org/#1}{\path{#1}}}
\providecommand{\Pubmed}[1]{\href{pmid:#1}{\path{#1}}}
\providecommand{\bibinfo}[2]{#2}
\ifx\xfnm\relax \def\xfnm[#1]{\unskip,\space#1}\fi
%Type = Article
\bibitem[{Adadi and Berrada(2018)}]{adadi2018peeking}
\bibinfo{author}{Adadi, A.}, \bibinfo{author}{Berrada, M.},
  \bibinfo{year}{2018}.
\newblock \bibinfo{title}{Peeking inside the black-box: a survey on explainable
  artificial intelligence (xai)}.
\newblock \bibinfo{journal}{IEEE access} \bibinfo{volume}{6},
  \bibinfo{pages}{52138--52160}.
%Type = Article
\bibitem[{Aiello et~al.(2020)Aiello, Renson and Zivich}]{aiello2020social}
\bibinfo{author}{Aiello, A.E.}, \bibinfo{author}{Renson, A.},
  \bibinfo{author}{Zivich, P.}, \bibinfo{year}{2020}.
\newblock \bibinfo{title}{Social media-and internet-based disease surveillance
  for public health}.
\newblock \bibinfo{journal}{Annual review of public health}
  \bibinfo{volume}{41}, \bibinfo{pages}{101}.
%Type = Article
\bibitem[{Amann et~al.(2020)Amann, Blasimme, Vayena, Frey, Madai and
  Consortium}]{amann2020explainability}
\bibinfo{author}{Amann, J.}, \bibinfo{author}{Blasimme, A.},
  \bibinfo{author}{Vayena, E.}, \bibinfo{author}{Frey, D.},
  \bibinfo{author}{Madai, V.I.}, \bibinfo{author}{Consortium, P.},
  \bibinfo{year}{2020}.
\newblock \bibinfo{title}{Explainability for artificial intelligence in
  healthcare: a multidisciplinary perspective}.
\newblock \bibinfo{journal}{BMC medical informatics and decision making}
  \bibinfo{volume}{20}, \bibinfo{pages}{1--9}.
%Type = Inproceedings
\bibitem[{Arnold and Kasenberg(2017)}]{arnold2017value}
\bibinfo{author}{Arnold, T.}, \bibinfo{author}{Kasenberg, D.},
  \bibinfo{year}{2017}.
\newblock \bibinfo{title}{Value alignment or misalignment {\^a}€“what will
  keep systems accountable?}, in: \bibinfo{booktitle}{AAAI Workshop on AI,
  Ethics, and Society}.
%Type = Article
\bibitem[{Arrieta et~al.(2020)Arrieta, D{\'\i}az-Rodr{\'\i}guez, Del~Ser,
  Bennetot, Tabik, Barbado, Garc{\'\i}a, Gil-L{\'o}pez, Molina, Benjamins
  et~al.}]{arrieta2020explainable}
\bibinfo{author}{Arrieta, A.B.}, \bibinfo{author}{D{\'\i}az-Rodr{\'\i}guez,
  N.}, \bibinfo{author}{Del~Ser, J.}, \bibinfo{author}{Bennetot, A.},
  \bibinfo{author}{Tabik, S.}, \bibinfo{author}{Barbado, A.},
  \bibinfo{author}{Garc{\'\i}a, S.}, \bibinfo{author}{Gil-L{\'o}pez, S.},
  \bibinfo{author}{Molina, D.}, \bibinfo{author}{Benjamins, R.}, et~al.,
  \bibinfo{year}{2020}.
\newblock \bibinfo{title}{Explainable artificial intelligence (xai): Concepts,
  taxonomies, opportunities and challenges toward responsible ai}.
\newblock \bibinfo{journal}{Information fusion} \bibinfo{volume}{58},
  \bibinfo{pages}{82--115}.
%Type = Misc
\bibitem[{Asuncion and Newman(2007)}]{asuncion2007uci}
\bibinfo{author}{Asuncion, A.}, \bibinfo{author}{Newman, D.},
  \bibinfo{year}{2007}.
\newblock \bibinfo{title}{Uci machine learning repository}.
%Type = Article
\bibitem[{Attard-Frost et~al.(2022)Attard-Frost, De~los R{\'\i}os and
  Walters}]{attard2022ethics}
\bibinfo{author}{Attard-Frost, B.}, \bibinfo{author}{De~los R{\'\i}os, A.},
  \bibinfo{author}{Walters, D.R.}, \bibinfo{year}{2022}.
\newblock \bibinfo{title}{The ethics of ai business practices: a review of 47
  ai ethics guidelines}.
\newblock \bibinfo{journal}{AI and Ethics} , \bibinfo{pages}{1--18}.
%Type = Article
\bibitem[{Azeroual et~al.(2018)Azeroual, Saake and
  Schallehn}]{azeroual2018analyzing}
\bibinfo{author}{Azeroual, O.}, \bibinfo{author}{Saake, G.},
  \bibinfo{author}{Schallehn, E.}, \bibinfo{year}{2018}.
\newblock \bibinfo{title}{Analyzing data quality issues in research information
  systems via data profiling}.
\newblock \bibinfo{journal}{International Journal of Information Management}
  \bibinfo{volume}{41}, \bibinfo{pages}{50--56}.
%Type = Article
\bibitem[{Bear Don’t Walk~IV et~al.(2022)Bear Don’t Walk~IV, Reyes~Nieva,
  Lee and Elhadad}]{bear2022scoping}
\bibinfo{author}{Bear Don’t Walk~IV, O.J.}, \bibinfo{author}{Reyes~Nieva,
  H.}, \bibinfo{author}{Lee, S.S.J.}, \bibinfo{author}{Elhadad, N.},
  \bibinfo{year}{2022}.
\newblock \bibinfo{title}{A scoping review of ethics considerations in clinical
  natural language processing}.
\newblock \bibinfo{journal}{JAMIA open} \bibinfo{volume}{5},
  \bibinfo{pages}{ooac039}.
%Type = Article
\bibitem[{Bellamy et~al.(2019)Bellamy, Dey, Hind, Hoffman, Houde, Kannan,
  Lohia, Martino, Mehta, Mojsilovi{\'c} et~al.}]{bellamy2019ai}
\bibinfo{author}{Bellamy, R.K.}, \bibinfo{author}{Dey, K.},
  \bibinfo{author}{Hind, M.}, \bibinfo{author}{Hoffman, S.C.},
  \bibinfo{author}{Houde, S.}, \bibinfo{author}{Kannan, K.},
  \bibinfo{author}{Lohia, P.}, \bibinfo{author}{Martino, J.},
  \bibinfo{author}{Mehta, S.}, \bibinfo{author}{Mojsilovi{\'c}, A.}, et~al.,
  \bibinfo{year}{2019}.
\newblock \bibinfo{title}{Ai fairness 360: An extensible toolkit for detecting
  and mitigating algorithmic bias}.
\newblock \bibinfo{journal}{IBM Journal of Research and Development}
  \bibinfo{volume}{63}, \bibinfo{pages}{4--1}.
%Type = Article
\bibitem[{Bertino et~al.(2019)Bertino, Merrill, Nesen and
  Utz}]{bertino2019redefining}
\bibinfo{author}{Bertino, E.}, \bibinfo{author}{Merrill, S.},
  \bibinfo{author}{Nesen, A.}, \bibinfo{author}{Utz, C.}, \bibinfo{year}{2019}.
\newblock \bibinfo{title}{Redefining data transparency: A multidimensional
  approach}.
\newblock \bibinfo{journal}{Computer} \bibinfo{volume}{52},
  \bibinfo{pages}{16--26}.
%Type = Article
\bibitem[{Bhatia-Lin et~al.(2019)Bhatia-Lin, Boon-Dooley, Roberts, Pronai,
  Fisher, Parker, Engstrom, Ingraham and Darnell}]{bhatia2019ethical}
\bibinfo{author}{Bhatia-Lin, A.}, \bibinfo{author}{Boon-Dooley, A.},
  \bibinfo{author}{Roberts, M.K.}, \bibinfo{author}{Pronai, C.},
  \bibinfo{author}{Fisher, D.}, \bibinfo{author}{Parker, L.},
  \bibinfo{author}{Engstrom, A.}, \bibinfo{author}{Ingraham, L.},
  \bibinfo{author}{Darnell, D.}, \bibinfo{year}{2019}.
\newblock \bibinfo{title}{Ethical and regulatory considerations for using
  social media platforms to locate and track research participants}.
\newblock \bibinfo{journal}{The American Journal of Bioethics}
  \bibinfo{volume}{19}, \bibinfo{pages}{47--61}.
%Type = Article
\bibitem[{Bird et~al.(2020)Bird, Dud{\'\i}k, Edgar, Horn, Lutz, Milan, Sameki,
  Wallach and Walker}]{bird2020fairlearn}
\bibinfo{author}{Bird, S.}, \bibinfo{author}{Dud{\'\i}k, M.},
  \bibinfo{author}{Edgar, R.}, \bibinfo{author}{Horn, B.},
  \bibinfo{author}{Lutz, R.}, \bibinfo{author}{Milan, V.},
  \bibinfo{author}{Sameki, M.}, \bibinfo{author}{Wallach, H.},
  \bibinfo{author}{Walker, K.}, \bibinfo{year}{2020}.
\newblock \bibinfo{title}{Fairlearn: A toolkit for assessing and improving
  fairness in ai}.
\newblock \bibinfo{journal}{Microsoft, Tech. Rep. MSR-TR-2020-32} .
%Type = Article
\bibitem[{Blacklaws(2018)}]{blacklaws2018algorithms}
\bibinfo{author}{Blacklaws, C.}, \bibinfo{year}{2018}.
\newblock \bibinfo{title}{Algorithms: transparency and accountability}.
\newblock \bibinfo{journal}{Philosophical Transactions of the Royal Society A:
  Mathematical, Physical and Engineering Sciences} \bibinfo{volume}{376},
  \bibinfo{pages}{20170351}.
%Type = Article
\bibitem[{Bose and Frew(2005)}]{bose2005lineage}
\bibinfo{author}{Bose, R.}, \bibinfo{author}{Frew, J.}, \bibinfo{year}{2005}.
\newblock \bibinfo{title}{Lineage retrieval for scientific data processing: a
  survey}.
\newblock \bibinfo{journal}{ACM Computing Surveys (CSUR)} \bibinfo{volume}{37},
  \bibinfo{pages}{1--28}.
%Type = Article
\bibitem[{Brundage et~al.(2020)Brundage, Avin, Wang, Belfield, Krueger,
  Hadfield, Khlaaf, Yang, Toner, Fong et~al.}]{brundage2020toward}
\bibinfo{author}{Brundage, M.}, \bibinfo{author}{Avin, S.},
  \bibinfo{author}{Wang, J.}, \bibinfo{author}{Belfield, H.},
  \bibinfo{author}{Krueger, G.}, \bibinfo{author}{Hadfield, G.},
  \bibinfo{author}{Khlaaf, H.}, \bibinfo{author}{Yang, J.},
  \bibinfo{author}{Toner, H.}, \bibinfo{author}{Fong, R.}, et~al.,
  \bibinfo{year}{2020}.
\newblock \bibinfo{title}{Toward trustworthy ai development: mechanisms for
  supporting verifiable claims}.
\newblock \bibinfo{journal}{arXiv preprint arXiv:2004.07213} .
%Type = Inproceedings
\bibitem[{Bucher et~al.(2016)Bucher, Herbin and Jurie}]{bucher2016improving}
\bibinfo{author}{Bucher, M.}, \bibinfo{author}{Herbin, S.},
  \bibinfo{author}{Jurie, F.}, \bibinfo{year}{2016}.
\newblock \bibinfo{title}{Improving semantic embedding consistency by metric
  learning for zero-shot classiffication}, in: \bibinfo{booktitle}{Computer
  Vision--ECCV 2016: 14th European Conference, Amsterdam, The Netherlands,
  October 11-14, 2016, Proceedings, Part V 14},
  \bibinfo{organization}{Springer}. pp. \bibinfo{pages}{730--746}.
%Type = Inproceedings
\bibitem[{Burgess et~al.(2022)Burgess, Hart, Elsayed, Cerny, Bures and
  Tisnovsky}]{burgess2022visualizing}
\bibinfo{author}{Burgess, K.}, \bibinfo{author}{Hart, D.},
  \bibinfo{author}{Elsayed, A.}, \bibinfo{author}{Cerny, T.},
  \bibinfo{author}{Bures, M.}, \bibinfo{author}{Tisnovsky, P.},
  \bibinfo{year}{2022}.
\newblock \bibinfo{title}{Visualizing architectural evolution via provenance
  tracking: a systematic review}, in: \bibinfo{booktitle}{Proceedings of the
  Conference on Research in Adaptive and Convergent Systems}, pp.
  \bibinfo{pages}{83--91}.
%Type = Article
\bibitem[{Carvalho et~al.(2019)Carvalho, Pereira and
  Cardoso}]{carvalho2019machine}
\bibinfo{author}{Carvalho, D.V.}, \bibinfo{author}{Pereira, E.M.},
  \bibinfo{author}{Cardoso, J.S.}, \bibinfo{year}{2019}.
\newblock \bibinfo{title}{Machine learning interpretability: A survey on
  methods and metrics}.
\newblock \bibinfo{journal}{Electronics} \bibinfo{volume}{8},
  \bibinfo{pages}{832}.
%Type = Article
\bibitem[{Chakraborti et~al.(2020)Chakraborti, Patra and
  Noble}]{chakraborti2020contrastive}
\bibinfo{author}{Chakraborti, T.}, \bibinfo{author}{Patra, A.},
  \bibinfo{author}{Noble, J.A.}, \bibinfo{year}{2020}.
\newblock \bibinfo{title}{Contrastive fairness in machine learning}.
\newblock \bibinfo{journal}{IEEE Letters of the Computer Society}
  \bibinfo{volume}{3}, \bibinfo{pages}{38--41}.
%Type = Inproceedings
\bibitem[{Chakraborty et~al.(2017)Chakraborty, Tomsett, Raghavendra, Harborne,
  Alzantot, Cerutti, Srivastava, Preece, Julier, Rao
  et~al.}]{chakraborty2017interpretability}
\bibinfo{author}{Chakraborty, S.}, \bibinfo{author}{Tomsett, R.},
  \bibinfo{author}{Raghavendra, R.}, \bibinfo{author}{Harborne, D.},
  \bibinfo{author}{Alzantot, M.}, \bibinfo{author}{Cerutti, F.},
  \bibinfo{author}{Srivastava, M.}, \bibinfo{author}{Preece, A.},
  \bibinfo{author}{Julier, S.}, \bibinfo{author}{Rao, R.M.}, et~al.,
  \bibinfo{year}{2017}.
\newblock \bibinfo{title}{Interpretability of deep learning models: A survey of
  results}, in: \bibinfo{booktitle}{2017 IEEE smartworld, ubiquitous
  intelligence \& computing, advanced \& trusted computed, scalable computing
  \& communications, cloud \& big data computing, Internet of people and smart
  city innovation (smartworld/SCALCOM/UIC/ATC/CBDcom/IOP/SCI)},
  \bibinfo{organization}{IEEE}. pp. \bibinfo{pages}{1--6}.
%Type = Article
\bibitem[{Chen and Wang(2021)}]{chen2021social}
\bibinfo{author}{Chen, J.}, \bibinfo{author}{Wang, Y.}, \bibinfo{year}{2021}.
\newblock \bibinfo{title}{Social media use for health purposes: systematic
  review}.
\newblock \bibinfo{journal}{Journal of medical Internet research}
  \bibinfo{volume}{23}, \bibinfo{pages}{e17917}.
%Type = Article
\bibitem[{Chouldechova and Roth(2020)}]{chouldechova2020snapshot}
\bibinfo{author}{Chouldechova, A.}, \bibinfo{author}{Roth, A.},
  \bibinfo{year}{2020}.
\newblock \bibinfo{title}{A snapshot of the frontiers of fairness in machine
  learning}.
\newblock \bibinfo{journal}{Communications of the ACM} \bibinfo{volume}{63},
  \bibinfo{pages}{82--89}.
%Type = Article
\bibitem[{Conway and O’Connor(2016)}]{conway2016social}
\bibinfo{author}{Conway, M.}, \bibinfo{author}{O’Connor, D.},
  \bibinfo{year}{2016}.
\newblock \bibinfo{title}{Social media, big data, and mental health: current
  advances and ethical implications}.
\newblock \bibinfo{journal}{Current opinion in psychology} \bibinfo{volume}{9},
  \bibinfo{pages}{77--82}.
%Type = Article
\bibitem[{Crawley et~al.(2021)Crawley, Divi and Smolinski}]{crawley2021using}
\bibinfo{author}{Crawley, A.W.}, \bibinfo{author}{Divi, N.},
  \bibinfo{author}{Smolinski, M.S.}, \bibinfo{year}{2021}.
\newblock \bibinfo{title}{Using timeliness metrics to track progress and
  identify gaps in disease surveillance}.
\newblock \bibinfo{journal}{Health security} \bibinfo{volume}{19},
  \bibinfo{pages}{309--317}.
%Type = Article
\bibitem[{Crockett(2013)}]{crockett2013models}
\bibinfo{author}{Crockett, M.J.}, \bibinfo{year}{2013}.
\newblock \bibinfo{title}{Models of morality}.
\newblock \bibinfo{journal}{Trends in cognitive sciences} \bibinfo{volume}{17},
  \bibinfo{pages}{363--366}.
%Type = Inproceedings
\bibitem[{Datta et~al.(2016)Datta, Sen and Zick}]{datta2016algorithmic}
\bibinfo{author}{Datta, A.}, \bibinfo{author}{Sen, S.}, \bibinfo{author}{Zick,
  Y.}, \bibinfo{year}{2016}.
\newblock \bibinfo{title}{Algorithmic transparency via quantitative input
  influence: Theory and experiments with learning systems}, in:
  \bibinfo{booktitle}{2016 IEEE symposium on security and privacy (SP)},
  \bibinfo{organization}{IEEE}. pp. \bibinfo{pages}{598--617}.
%Type = Book
\bibitem[{Davis(2012)}]{davis2012ethics}
\bibinfo{author}{Davis, K.}, \bibinfo{year}{2012}.
\newblock \bibinfo{title}{Ethics of Big Data: Balancing risk and innovation}.
\newblock \bibinfo{publisher}{" O'Reilly Media, Inc."}.
%Type = Article
\bibitem[{Diakopoulos and Koliska(2017)}]{diakopoulos2017algorithmic}
\bibinfo{author}{Diakopoulos, N.}, \bibinfo{author}{Koliska, M.},
  \bibinfo{year}{2017}.
\newblock \bibinfo{title}{Algorithmic transparency in the news media}.
\newblock \bibinfo{journal}{Digital journalism} \bibinfo{volume}{5},
  \bibinfo{pages}{809--828}.
%Type = Inproceedings
\bibitem[{Dixon et~al.(2018)Dixon, Li, Sorensen, Thain and
  Vasserman}]{dixon2018measuring}
\bibinfo{author}{Dixon, L.}, \bibinfo{author}{Li, J.},
  \bibinfo{author}{Sorensen, J.}, \bibinfo{author}{Thain, N.},
  \bibinfo{author}{Vasserman, L.}, \bibinfo{year}{2018}.
\newblock \bibinfo{title}{Measuring and mitigating unintended bias in text
  classification}, in: \bibinfo{booktitle}{Proceedings of the 2018 AAAI/ACM
  Conference on AI, Ethics, and Society}, pp. \bibinfo{pages}{67--73}.
%Type = Article
\bibitem[{Drabiak and Wolfson(2020)}]{drabiak2020should}
\bibinfo{author}{Drabiak, K.}, \bibinfo{author}{Wolfson, J.},
  \bibinfo{year}{2020}.
\newblock \bibinfo{title}{What should health care organizations do to reduce
  billing fraud and abuse?}
\newblock \bibinfo{journal}{AMA Journal of Ethics} \bibinfo{volume}{22},
  \bibinfo{pages}{221--231}.
%Type = Article
\bibitem[{Dua et~al.(2017)Dua, Graff et~al.}]{dua2017uci}
\bibinfo{author}{Dua, D.}, \bibinfo{author}{Graff, C.}, et~al.,
  \bibinfo{year}{2017}.
\newblock \bibinfo{title}{Uci machine learning repository, 2017}.
\newblock \bibinfo{journal}{URL http://archive. ics. uci. edu/ml}
  \bibinfo{volume}{7}.
%Type = Book
\bibitem[{Dubberley et~al.(2020)Dubberley, Koenig and
  Murray}]{dubberley2020digital}
\bibinfo{author}{Dubberley, S.}, \bibinfo{author}{Koenig, A.},
  \bibinfo{author}{Murray, D.}, \bibinfo{year}{2020}.
\newblock \bibinfo{title}{Digital witness: using open source information for
  human rights investigation, documentation, and accountability}.
\newblock \bibinfo{publisher}{Oxford University Press, USA}.
%Type = Article
\bibitem[{Enarsson et~al.(2022)Enarsson, Enqvist and
  Naarttij{\"a}rvi}]{enarsson2022approaching}
\bibinfo{author}{Enarsson, T.}, \bibinfo{author}{Enqvist, L.},
  \bibinfo{author}{Naarttij{\"a}rvi, M.}, \bibinfo{year}{2022}.
\newblock \bibinfo{title}{Approaching the human in the loop--legal perspectives
  on hybrid human/algorithmic decision-making in three contexts}.
\newblock \bibinfo{journal}{Information \& Communications Technology Law}
  \bibinfo{volume}{31}, \bibinfo{pages}{123--153}.
%Type = Article
\bibitem[{Fan et~al.(2011)Fan, Chang, Lin and Hsieh}]{fan2011hybrid}
\bibinfo{author}{Fan, C.Y.}, \bibinfo{author}{Chang, P.C.},
  \bibinfo{author}{Lin, J.J.}, \bibinfo{author}{Hsieh, J.},
  \bibinfo{year}{2011}.
\newblock \bibinfo{title}{A hybrid model combining case-based reasoning and
  fuzzy decision tree for medical data classification}.
\newblock \bibinfo{journal}{Applied Soft Computing} \bibinfo{volume}{11},
  \bibinfo{pages}{632--644}.
%Type = Article
\bibitem[{Gagnon and Sabus(2015)}]{gagnon2015professionalism}
\bibinfo{author}{Gagnon, K.}, \bibinfo{author}{Sabus, C.},
  \bibinfo{year}{2015}.
\newblock \bibinfo{title}{Professionalism in a digital age: opportunities and
  considerations for using social media in health care}.
\newblock \bibinfo{journal}{Physical therapy} \bibinfo{volume}{95},
  \bibinfo{pages}{406--414}.
%Type = Inproceedings
\bibitem[{Ghassami et~al.(2018)Ghassami, Khodadadian and
  Kiyavash}]{ghassami2018fairness}
\bibinfo{author}{Ghassami, A.}, \bibinfo{author}{Khodadadian, S.},
  \bibinfo{author}{Kiyavash, N.}, \bibinfo{year}{2018}.
\newblock \bibinfo{title}{Fairness in supervised learning: An information
  theoretic approach}, in: \bibinfo{booktitle}{2018 IEEE international
  symposium on information theory (ISIT)}, \bibinfo{organization}{IEEE}. pp.
  \bibinfo{pages}{176--180}.
%Type = Inproceedings
\bibitem[{Ghosh et~al.(2021)Ghosh, Genuit and Reagan}]{ghosh2021characterizing}
\bibinfo{author}{Ghosh, A.}, \bibinfo{author}{Genuit, L.},
  \bibinfo{author}{Reagan, M.}, \bibinfo{year}{2021}.
\newblock \bibinfo{title}{Characterizing intersectional group fairness with
  worst-case comparisons}, in: \bibinfo{booktitle}{Artificial Intelligence
  Diversity, Belonging, Equity, and Inclusion}, \bibinfo{organization}{PMLR}.
  pp. \bibinfo{pages}{22--34}.
%Type = Inproceedings
\bibitem[{Gilpin et~al.(2018)Gilpin, Bau, Yuan, Bajwa, Specter and
  Kagal}]{gilpin2018explaining}
\bibinfo{author}{Gilpin, L.H.}, \bibinfo{author}{Bau, D.},
  \bibinfo{author}{Yuan, B.Z.}, \bibinfo{author}{Bajwa, A.},
  \bibinfo{author}{Specter, M.}, \bibinfo{author}{Kagal, L.},
  \bibinfo{year}{2018}.
\newblock \bibinfo{title}{Explaining explanations: An overview of
  interpretability of machine learning}, in: \bibinfo{booktitle}{2018 IEEE 5th
  International Conference on data science and advanced analytics (DSAA)},
  \bibinfo{organization}{IEEE}. pp. \bibinfo{pages}{80--89}.
%Type = Article
\bibitem[{Golder et~al.(2017)Golder, Ahmed, Norman and
  Booth}]{golder2017attitudes}
\bibinfo{author}{Golder, S.}, \bibinfo{author}{Ahmed, S.},
  \bibinfo{author}{Norman, G.}, \bibinfo{author}{Booth, A.},
  \bibinfo{year}{2017}.
\newblock \bibinfo{title}{Attitudes toward the ethics of research using social
  media: a systematic review}.
\newblock \bibinfo{journal}{Journal of medical internet research}
  \bibinfo{volume}{19}, \bibinfo{pages}{e195}.
%Type = Article
\bibitem[{G{\'o}mez-Gonz{\'a}lez et~al.(2020)G{\'o}mez-Gonz{\'a}lez, Gomez,
  M{\'a}rquez-Rivas, Guerrero-Claro, Fern{\'a}ndez-Lizaranzu,
  Relimpio-L{\'o}pez, Dorado, Mayorga-Buiza, Izquierdo-Ayuso and
  Capit{\'a}n-Morales}]{gomez2020artificial}
\bibinfo{author}{G{\'o}mez-Gonz{\'a}lez, E.}, \bibinfo{author}{Gomez, E.},
  \bibinfo{author}{M{\'a}rquez-Rivas, J.}, \bibinfo{author}{Guerrero-Claro,
  M.}, \bibinfo{author}{Fern{\'a}ndez-Lizaranzu, I.},
  \bibinfo{author}{Relimpio-L{\'o}pez, M.I.}, \bibinfo{author}{Dorado, M.E.},
  \bibinfo{author}{Mayorga-Buiza, M.J.}, \bibinfo{author}{Izquierdo-Ayuso, G.},
  \bibinfo{author}{Capit{\'a}n-Morales, L.}, \bibinfo{year}{2020}.
\newblock \bibinfo{title}{Artificial intelligence in medicine and healthcare: a
  review and classification of current and near-future applications and their
  ethical and social impact}.
\newblock \bibinfo{journal}{arXiv preprint arXiv:2001.09778} .
%Type = Article
\bibitem[{Grajales~III et~al.(2014)Grajales~III, Sheps, Ho, Novak-Lauscher and
  Eysenbach}]{grajales2014social}
\bibinfo{author}{Grajales~III, F.J.}, \bibinfo{author}{Sheps, S.},
  \bibinfo{author}{Ho, K.}, \bibinfo{author}{Novak-Lauscher, H.},
  \bibinfo{author}{Eysenbach, G.}, \bibinfo{year}{2014}.
\newblock \bibinfo{title}{Social media: a review and tutorial of applications
  in medicine and health care}.
\newblock \bibinfo{journal}{Journal of medical Internet research}
  \bibinfo{volume}{16}, \bibinfo{pages}{e2912}.
%Type = Article
\bibitem[{Grandini et~al.(2020)Grandini, Bagli and
  Visani}]{grandini2020metrics}
\bibinfo{author}{Grandini, M.}, \bibinfo{author}{Bagli, E.},
  \bibinfo{author}{Visani, G.}, \bibinfo{year}{2020}.
\newblock \bibinfo{title}{Metrics for multi-class classification: an overview}.
\newblock \bibinfo{journal}{arXiv preprint arXiv:2008.05756} .
%Type = Article
\bibitem[{Gu et~al.(2022)Gu, Gao, Chen, Shi, Li and Cao}]{gu2022electricity}
\bibinfo{author}{Gu, D.}, \bibinfo{author}{Gao, Y.}, \bibinfo{author}{Chen,
  K.}, \bibinfo{author}{Shi, J.}, \bibinfo{author}{Li, Y.},
  \bibinfo{author}{Cao, Y.}, \bibinfo{year}{2022}.
\newblock \bibinfo{title}{Electricity theft detection in ami with low false
  positive rate based on deep learning and evolutionary algorithm}.
\newblock \bibinfo{journal}{IEEE Transactions on Power Systems}
  \bibinfo{volume}{37}, \bibinfo{pages}{4568--4578}.
%Type = Incollection
\bibitem[{Guttman(2017)}]{guttman2017ethical}
\bibinfo{author}{Guttman, N.}, \bibinfo{year}{2017}.
\newblock \bibinfo{title}{Ethical issues in health promotion and communication
  interventions}, in: \bibinfo{booktitle}{Oxford research encyclopedia of
  communication}.
%Type = Article
\bibitem[{Hagerty and Rubinov(2019)}]{hagerty2019global}
\bibinfo{author}{Hagerty, A.}, \bibinfo{author}{Rubinov, I.},
  \bibinfo{year}{2019}.
\newblock \bibinfo{title}{Global ai ethics: a review of the social impacts and
  ethical implications of artificial intelligence}.
\newblock \bibinfo{journal}{arXiv preprint arXiv:1907.07892} .
%Type = Article
\bibitem[{Hansen(1997)}]{hansen1997hipaa}
\bibinfo{author}{Hansen, E.}, \bibinfo{year}{1997}.
\newblock \bibinfo{title}{Hipaa (health insurance portability and
  accountability act) rules: federal and state enforcement.}
\newblock \bibinfo{journal}{Medical Interface} \bibinfo{volume}{10},
  \bibinfo{pages}{96--8}.
%Type = Article
\bibitem[{He et~al.(2019)He, Baxter, Xu, Xu, Zhou and Zhang}]{he2019practical}
\bibinfo{author}{He, J.}, \bibinfo{author}{Baxter, S.L.}, \bibinfo{author}{Xu,
  J.}, \bibinfo{author}{Xu, J.}, \bibinfo{author}{Zhou, X.},
  \bibinfo{author}{Zhang, K.}, \bibinfo{year}{2019}.
\newblock \bibinfo{title}{The practical implementation of artificial
  intelligence technologies in medicine}.
\newblock \bibinfo{journal}{Nature medicine} \bibinfo{volume}{25},
  \bibinfo{pages}{30--36}.
%Type = Inproceedings
\bibitem[{Hertweck et~al.(2021)Hertweck, Heitz and Loi}]{hertweck2021moral}
\bibinfo{author}{Hertweck, C.}, \bibinfo{author}{Heitz, C.},
  \bibinfo{author}{Loi, M.}, \bibinfo{year}{2021}.
\newblock \bibinfo{title}{On the moral justification of statistical parity},
  in: \bibinfo{booktitle}{Proceedings of the 2021 ACM Conference on Fairness,
  Accountability, and Transparency}, pp. \bibinfo{pages}{747--757}.
%Type = Inproceedings
\bibitem[{Hossin et~al.(2011)Hossin, Sulaiman, Mustapha, Mustapha and
  Rahmat}]{hossin2011hybrid}
\bibinfo{author}{Hossin, M.}, \bibinfo{author}{Sulaiman, M.},
  \bibinfo{author}{Mustapha, A.}, \bibinfo{author}{Mustapha, N.},
  \bibinfo{author}{Rahmat, R.}, \bibinfo{year}{2011}.
\newblock \bibinfo{title}{A hybrid evaluation metric for optimizing
  classifier}, in: \bibinfo{booktitle}{2011 3rd Conference on Data Mining and
  Optimization (DMO)}, \bibinfo{organization}{IEEE}. pp.
  \bibinfo{pages}{165--170}.
%Type = Inproceedings
\bibitem[{Hutchinson et~al.(2021)Hutchinson, Smart, Hanna, Denton, Greer,
  Kjartansson, Barnes and Mitchell}]{hutchinson2021towards}
\bibinfo{author}{Hutchinson, B.}, \bibinfo{author}{Smart, A.},
  \bibinfo{author}{Hanna, A.}, \bibinfo{author}{Denton, E.},
  \bibinfo{author}{Greer, C.}, \bibinfo{author}{Kjartansson, O.},
  \bibinfo{author}{Barnes, P.}, \bibinfo{author}{Mitchell, M.},
  \bibinfo{year}{2021}.
\newblock \bibinfo{title}{Towards accountability for machine learning datasets:
  Practices from software engineering and infrastructure}, in:
  \bibinfo{booktitle}{Proceedings of the 2021 ACM Conference on Fairness,
  Accountability, and Transparency}, pp. \bibinfo{pages}{560--575}.
%Type = Inproceedings
\bibitem[{Iyer et~al.(2018)Iyer, Li, Li, Lewis, Sundar and
  Sycara}]{iyer2018transparency}
\bibinfo{author}{Iyer, R.}, \bibinfo{author}{Li, Y.}, \bibinfo{author}{Li, H.},
  \bibinfo{author}{Lewis, M.}, \bibinfo{author}{Sundar, R.},
  \bibinfo{author}{Sycara, K.}, \bibinfo{year}{2018}.
\newblock \bibinfo{title}{Transparency and explanation in deep reinforcement
  learning neural networks}, in: \bibinfo{booktitle}{Proceedings of the 2018
  AAAI/ACM Conference on AI, Ethics, and Society}, pp.
  \bibinfo{pages}{144--150}.
%Type = Inproceedings
\bibitem[{Jadon(2020)}]{jadon2020survey}
\bibinfo{author}{Jadon, S.}, \bibinfo{year}{2020}.
\newblock \bibinfo{title}{A survey of loss functions for semantic
  segmentation}, in: \bibinfo{booktitle}{2020 IEEE conference on computational
  intelligence in bioinformatics and computational biology (CIBCB)},
  \bibinfo{organization}{IEEE}. pp. \bibinfo{pages}{1--7}.
%Type = Inproceedings
\bibitem[{Jakesch et~al.(2022)Jakesch, Bu{\c{c}}inca, Amershi and
  Olteanu}]{jakesch2022different}
\bibinfo{author}{Jakesch, M.}, \bibinfo{author}{Bu{\c{c}}inca, Z.},
  \bibinfo{author}{Amershi, S.}, \bibinfo{author}{Olteanu, A.},
  \bibinfo{year}{2022}.
\newblock \bibinfo{title}{How different groups prioritize ethical values for
  responsible ai}, in: \bibinfo{booktitle}{2022 ACM Conference on Fairness,
  Accountability, and Transparency}, pp. \bibinfo{pages}{310--323}.
%Type = Article
\bibitem[{Janssen et~al.(2022)Janssen, Hartog, Matheus, Yi~Ding and
  Kuk}]{janssen2022will}
\bibinfo{author}{Janssen, M.}, \bibinfo{author}{Hartog, M.},
  \bibinfo{author}{Matheus, R.}, \bibinfo{author}{Yi~Ding, A.},
  \bibinfo{author}{Kuk, G.}, \bibinfo{year}{2022}.
\newblock \bibinfo{title}{Will algorithms blind people? the effect of
  explainable ai and decision-makers’ experience on ai-supported
  decision-making in government}.
\newblock \bibinfo{journal}{Social Science Computer Review}
  \bibinfo{volume}{40}, \bibinfo{pages}{478--493}.
%Type = Article
\bibitem[{Johnson(2019)}]{johnson2019ai}
\bibinfo{author}{Johnson, S.L.}, \bibinfo{year}{2019}.
\newblock \bibinfo{title}{Ai, machine learning, and ethics in health care}.
\newblock \bibinfo{journal}{Journal of Legal Medicine} \bibinfo{volume}{39},
  \bibinfo{pages}{427--441}.
%Type = Article
\bibitem[{Kalkman et~al.(2019)Kalkman, Mostert, Gerlinger, van Delden and van
  Thiel}]{kalkman2019responsible}
\bibinfo{author}{Kalkman, S.}, \bibinfo{author}{Mostert, M.},
  \bibinfo{author}{Gerlinger, C.}, \bibinfo{author}{van Delden, J.J.},
  \bibinfo{author}{van Thiel, G.J.}, \bibinfo{year}{2019}.
\newblock \bibinfo{title}{Responsible data sharing in international health
  research: a systematic review of principles and norms}.
\newblock \bibinfo{journal}{BMC medical ethics} \bibinfo{volume}{20},
  \bibinfo{pages}{1--13}.
%Type = Article
\bibitem[{Kass and Faden(2018)}]{kass2018ethics}
\bibinfo{author}{Kass, N.E.}, \bibinfo{author}{Faden, R.R.},
  \bibinfo{year}{2018}.
\newblock \bibinfo{title}{Ethics and learning health care: the essential roles
  of engagement, transparency, and accountability}.
\newblock \bibinfo{journal}{Learning Health Systems} \bibinfo{volume}{2},
  \bibinfo{pages}{e10066}.
%Type = Inproceedings
\bibitem[{Kaur et~al.(2021)Kaur, Uslu, Durresi, Badve and
  Dundar}]{kaur2021trustworthy}
\bibinfo{author}{Kaur, D.}, \bibinfo{author}{Uslu, S.},
  \bibinfo{author}{Durresi, A.}, \bibinfo{author}{Badve, S.},
  \bibinfo{author}{Dundar, M.}, \bibinfo{year}{2021}.
\newblock \bibinfo{title}{Trustworthy explainability acceptance: A new metric
  to measure the trustworthiness of interpretable ai medical diagnostic
  systems}, in: \bibinfo{booktitle}{Complex, Intelligent and Software Intensive
  Systems: Proceedings of the 15th International Conference on Complex,
  Intelligent and Software Intensive Systems (CISIS-2021)},
  \bibinfo{organization}{Springer}. pp. \bibinfo{pages}{35--46}.
%Type = Article
\bibitem[{Kazim and Koshiyama(2021)}]{kazim2021high}
\bibinfo{author}{Kazim, E.}, \bibinfo{author}{Koshiyama, A.S.},
  \bibinfo{year}{2021}.
\newblock \bibinfo{title}{A high-level overview of ai ethics}.
\newblock \bibinfo{journal}{Patterns} \bibinfo{volume}{2},
  \bibinfo{pages}{100314}.
%Type = Article
\bibitem[{Kerikm{\"a}e and P{\"a}rn-Lee(2021)}]{kerikmae2021legal}
\bibinfo{author}{Kerikm{\"a}e, T.}, \bibinfo{author}{P{\"a}rn-Lee, E.},
  \bibinfo{year}{2021}.
\newblock \bibinfo{title}{Legal dilemmas of estonian artificial intelligence
  strategy: in between of e-society and global race}.
\newblock \bibinfo{journal}{Ai \& Society} \bibinfo{volume}{36},
  \bibinfo{pages}{561--572}.
%Type = Article
\bibitem[{Keshk et~al.(2022)Keshk, Moustafa, Sitnikova and
  Turnbull}]{keshk2022privacy}
\bibinfo{author}{Keshk, M.}, \bibinfo{author}{Moustafa, N.},
  \bibinfo{author}{Sitnikova, E.}, \bibinfo{author}{Turnbull, B.},
  \bibinfo{year}{2022}.
\newblock \bibinfo{title}{Privacy-preserving big data analytics for
  cyber-physical systems}.
\newblock \bibinfo{journal}{Wireless Networks} , \bibinfo{pages}{1--9}.
%Type = Incollection
\bibitem[{King(2013)}]{king2013instrumental}
\bibinfo{author}{King, J.}, \bibinfo{year}{2013}.
\newblock \bibinfo{title}{The instrumental value of legal accountability},
  \bibinfo{publisher}{Oxford University Press}.
%Type = Article
\bibitem[{Kington et~al.(2021)Kington, Arnesen, Chou, Curry, Lazer and
  Villarruel}]{kington2021identifying}
\bibinfo{author}{Kington, R.S.}, \bibinfo{author}{Arnesen, S.},
  \bibinfo{author}{Chou, W.Y.S.}, \bibinfo{author}{Curry, S.J.},
  \bibinfo{author}{Lazer, D.}, \bibinfo{author}{Villarruel, A.M.},
  \bibinfo{year}{2021}.
\newblock \bibinfo{title}{Identifying credible sources of health information in
  social media: Principles and attributes}.
\newblock \bibinfo{journal}{NAM perspectives} \bibinfo{volume}{2021}.
%Type = Inproceedings
\bibitem[{Ko et~al.(2011)Ko, Kirchberg and Lee}]{ko2011system}
\bibinfo{author}{Ko, R.K.}, \bibinfo{author}{Kirchberg, M.},
  \bibinfo{author}{Lee, B.S.}, \bibinfo{year}{2011}.
\newblock \bibinfo{title}{From system-centric to data-centric
  logging-accountability, trust \& security in cloud computing}, in:
  \bibinfo{booktitle}{2011 Defense Science Research Conference and Expo (DSR)},
  \bibinfo{organization}{IEEE}. pp. \bibinfo{pages}{1--4}.
%Type = Article
\bibitem[{Kofod-Petersen(2012)}]{kofod2012structured}
\bibinfo{author}{Kofod-Petersen, A.}, \bibinfo{year}{2012}.
\newblock \bibinfo{title}{How to do a structured literature review in computer
  science}.
\newblock \bibinfo{journal}{Ver. 0.1} \bibinfo{volume}{1}.
%Type = Article
\bibitem[{Kynk{\"a}{\"a}nniemi et~al.(2019)Kynk{\"a}{\"a}nniemi, Karras, Laine,
  Lehtinen and Aila}]{kynkaanniemi2019improved}
\bibinfo{author}{Kynk{\"a}{\"a}nniemi, T.}, \bibinfo{author}{Karras, T.},
  \bibinfo{author}{Laine, S.}, \bibinfo{author}{Lehtinen, J.},
  \bibinfo{author}{Aila, T.}, \bibinfo{year}{2019}.
\newblock \bibinfo{title}{Improved precision and recall metric for assessing
  generative models}.
\newblock \bibinfo{journal}{Advances in Neural Information Processing Systems}
  \bibinfo{volume}{32}.
%Type = Article
\bibitem[{Lagioia et~al.(2022)Lagioia, Rovatti and
  Sartor}]{lagioia2022algorithmic}
\bibinfo{author}{Lagioia, F.}, \bibinfo{author}{Rovatti, R.},
  \bibinfo{author}{Sartor, G.}, \bibinfo{year}{2022}.
\newblock \bibinfo{title}{Algorithmic fairness through group parities? the case
  of compas-sapmoc}.
\newblock \bibinfo{journal}{AI \& SOCIETY} , \bibinfo{pages}{1--20}.
%Type = Article
\bibitem[{Latonero(2018)}]{latonero2018governing}
\bibinfo{author}{Latonero, M.}, \bibinfo{year}{2018}.
\newblock \bibinfo{title}{Governing artificial intelligence: Upholding human
  rights \& dignity} .
%Type = Inproceedings
\bibitem[{Leidner and Plachouras(2017)}]{leidner2017ethical}
\bibinfo{author}{Leidner, J.L.}, \bibinfo{author}{Plachouras, V.},
  \bibinfo{year}{2017}.
\newblock \bibinfo{title}{Ethical by design: Ethics best practices for natural
  language processing}, in: \bibinfo{booktitle}{Proceedings of the First ACL
  Workshop on Ethics in Natural Language Processing}, pp.
  \bibinfo{pages}{30--40}.
%Type = Article
\bibitem[{Leikas et~al.(2019)Leikas, Koivisto and Gotcheva}]{leikas2019ethical}
\bibinfo{author}{Leikas, J.}, \bibinfo{author}{Koivisto, R.},
  \bibinfo{author}{Gotcheva, N.}, \bibinfo{year}{2019}.
\newblock \bibinfo{title}{Ethical framework for designing autonomous
  intelligent systems}.
\newblock \bibinfo{journal}{Journal of Open Innovation: Technology, Market, and
  Complexity} \bibinfo{volume}{5}, \bibinfo{pages}{18}.
%Type = Article
\bibitem[{Leonelli et~al.(2021)Leonelli, Lovell, Wheeler, Fleming and
  Williams}]{leonelli2021fair}
\bibinfo{author}{Leonelli, S.}, \bibinfo{author}{Lovell, R.},
  \bibinfo{author}{Wheeler, B.W.}, \bibinfo{author}{Fleming, L.},
  \bibinfo{author}{Williams, H.}, \bibinfo{year}{2021}.
\newblock \bibinfo{title}{From fair data to fair data use: Methodological data
  fairness in health-related social media research}.
\newblock \bibinfo{journal}{Big Data \& Society} \bibinfo{volume}{8},
  \bibinfo{pages}{20539517211010310}.
%Type = Article
\bibitem[{Leslie(2019)}]{leslie2019understanding}
\bibinfo{author}{Leslie, D.}, \bibinfo{year}{2019}.
\newblock \bibinfo{title}{Understanding artificial intelligence ethics and
  safety}.
\newblock \bibinfo{journal}{arXiv preprint arXiv:1906.05684} .
%Type = Article
\bibitem[{Li and Li(2020)}]{li2020overview}
\bibinfo{author}{Li, Q.}, \bibinfo{author}{Li, Q.}, \bibinfo{year}{2020}.
\newblock \bibinfo{title}{Overview of data visualization}.
\newblock \bibinfo{journal}{Embodying Data: Chinese Aesthetics, Interactive
  Visualization and Gaming Technologies} , \bibinfo{pages}{17--47}.
%Type = Book
\bibitem[{Link and Scott(2012)}]{link2012public}
\bibinfo{author}{Link, A.N.}, \bibinfo{author}{Scott, J.T.},
  \bibinfo{year}{2012}.
\newblock \bibinfo{title}{Public accountability: Evaluating technology-based
  institutions}.
\newblock \bibinfo{publisher}{Springer Science \& Business Media}.
%Type = Article
\bibitem[{Lipton(2018)}]{lipton2018mythos}
\bibinfo{author}{Lipton, Z.C.}, \bibinfo{year}{2018}.
\newblock \bibinfo{title}{The mythos of model interpretability: In machine
  learning, the concept of interpretability is both important and slippery.}
\newblock \bibinfo{journal}{Queue} \bibinfo{volume}{16},
  \bibinfo{pages}{31--57}.
%Type = Article
\bibitem[{Livingston et~al.(2012)Livingston, Milne, Fang and
  Amari}]{livingston2012effectiveness}
\bibinfo{author}{Livingston, J.D.}, \bibinfo{author}{Milne, T.},
  \bibinfo{author}{Fang, M.L.}, \bibinfo{author}{Amari, E.},
  \bibinfo{year}{2012}.
\newblock \bibinfo{title}{The effectiveness of interventions for reducing
  stigma related to substance use disorders: a systematic review}.
\newblock \bibinfo{journal}{Addiction} \bibinfo{volume}{107},
  \bibinfo{pages}{39--50}.
%Type = Article
\bibitem[{Lundberg and Lee(2017)}]{lundberg2017unified}
\bibinfo{author}{Lundberg, S.M.}, \bibinfo{author}{Lee, S.I.},
  \bibinfo{year}{2017}.
\newblock \bibinfo{title}{A unified approach to interpreting model
  predictions}.
\newblock \bibinfo{journal}{Advances in neural information processing systems}
  \bibinfo{volume}{30}.
%Type = Inproceedings
\bibitem[{Malawski(2020)}]{malawski2020note}
\bibinfo{author}{Malawski, M.}, \bibinfo{year}{2020}.
\newblock \bibinfo{title}{A note on equal treatment and symmetry of values},
  in: \bibinfo{booktitle}{Transactions on Computational Collective Intelligence
  XXXV}, \bibinfo{organization}{Springer}. pp. \bibinfo{pages}{76--84}.
%Type = Inproceedings
\bibitem[{Markoulidakis et~al.(2021)Markoulidakis, Kopsiaftis, Rallis and
  Georgoulas}]{markoulidakis2021multi}
\bibinfo{author}{Markoulidakis, I.}, \bibinfo{author}{Kopsiaftis, G.},
  \bibinfo{author}{Rallis, I.}, \bibinfo{author}{Georgoulas, I.},
  \bibinfo{year}{2021}.
\newblock \bibinfo{title}{Multi-class confusion matrix reduction method and its
  application on net promoter score classification problem}, in:
  \bibinfo{booktitle}{The 14th pervasive technologies related to assistive
  environments conference}, pp. \bibinfo{pages}{412--419}.
%Type = Inproceedings
\bibitem[{Mashhadi et~al.(2021)Mashhadi, Winder, Lia and Wood}]{mashhadi2021no}
\bibinfo{author}{Mashhadi, A.}, \bibinfo{author}{Winder, S.G.},
  \bibinfo{author}{Lia, E.H.}, \bibinfo{author}{Wood, S.A.},
  \bibinfo{year}{2021}.
\newblock \bibinfo{title}{No walk in the park: The viability and fairness of
  social media analysis for parks and recreational policy making}, in:
  \bibinfo{booktitle}{Proceedings of the International AAAI Conference on Web
  and Social Media}, pp. \bibinfo{pages}{409--420}.
%Type = Article
\bibitem[{Mehrabi et~al.(2021a)Mehrabi, Gupta, Morstatter, Steeg and
  Galstyan}]{mehrabi2021attributing}
\bibinfo{author}{Mehrabi, N.}, \bibinfo{author}{Gupta, U.},
  \bibinfo{author}{Morstatter, F.}, \bibinfo{author}{Steeg, G.V.},
  \bibinfo{author}{Galstyan, A.}, \bibinfo{year}{2021}a.
\newblock \bibinfo{title}{Attributing fair decisions with attention
  interventions}.
\newblock \bibinfo{journal}{arXiv preprint arXiv:2109.03952} .
%Type = Article
\bibitem[{Mehrabi et~al.(2021b)Mehrabi, Morstatter, Saxena, Lerman and
  Galstyan}]{mehrabi2021survey}
\bibinfo{author}{Mehrabi, N.}, \bibinfo{author}{Morstatter, F.},
  \bibinfo{author}{Saxena, N.}, \bibinfo{author}{Lerman, K.},
  \bibinfo{author}{Galstyan, A.}, \bibinfo{year}{2021}b.
\newblock \bibinfo{title}{A survey on bias and fairness in machine learning}.
\newblock \bibinfo{journal}{ACM Computing Surveys (CSUR)} \bibinfo{volume}{54},
  \bibinfo{pages}{1--35}.
%Type = Article
\bibitem[{Mendes and Vilela(2017)}]{mendes2017privacy}
\bibinfo{author}{Mendes, R.}, \bibinfo{author}{Vilela, J.P.},
  \bibinfo{year}{2017}.
\newblock \bibinfo{title}{Privacy-preserving data mining: methods, metrics, and
  applications}.
\newblock \bibinfo{journal}{IEEE Access} \bibinfo{volume}{5},
  \bibinfo{pages}{10562--10582}.
%Type = Article
\bibitem[{Mittelstadt(2019)}]{mittelstadt2019principles}
\bibinfo{author}{Mittelstadt, B.}, \bibinfo{year}{2019}.
\newblock \bibinfo{title}{Principles alone cannot guarantee ethical ai}.
\newblock \bibinfo{journal}{Nature machine intelligence} \bibinfo{volume}{1},
  \bibinfo{pages}{501--507}.
%Type = Article
\bibitem[{Murdoch et~al.(2019)Murdoch, Singh, Kumbier, Abbasi-Asl and
  Yu}]{murdoch2019definitions}
\bibinfo{author}{Murdoch, W.J.}, \bibinfo{author}{Singh, C.},
  \bibinfo{author}{Kumbier, K.}, \bibinfo{author}{Abbasi-Asl, R.},
  \bibinfo{author}{Yu, B.}, \bibinfo{year}{2019}.
\newblock \bibinfo{title}{Definitions, methods, and applications in
  interpretable machine learning}.
\newblock \bibinfo{journal}{Proceedings of the National Academy of Sciences}
  \bibinfo{volume}{116}, \bibinfo{pages}{22071--22080}.
%Type = Inproceedings
\bibitem[{Narasimhan et~al.(2020)Narasimhan, Cotter, Gupta and
  Wang}]{narasimhan2020pairwise}
\bibinfo{author}{Narasimhan, H.}, \bibinfo{author}{Cotter, A.},
  \bibinfo{author}{Gupta, M.}, \bibinfo{author}{Wang, S.},
  \bibinfo{year}{2020}.
\newblock \bibinfo{title}{Pairwise fairness for ranking and regression}, in:
  \bibinfo{booktitle}{Proceedings of the AAAI Conference on Artificial
  Intelligence}, pp. \bibinfo{pages}{5248--5255}.
%Type = Incollection
\bibitem[{Nebeker et~al.(2022)Nebeker, Parrish and Graham}]{nebeker2022ai}
\bibinfo{author}{Nebeker, C.}, \bibinfo{author}{Parrish, E.M.},
  \bibinfo{author}{Graham, S.}, \bibinfo{year}{2022}.
\newblock \bibinfo{title}{The ai artificial intelligence (ai)-powered digital
  health digital health sector: Ethical and regulatory considerations when
  developing digital mental health digital mental health mental health tools
  for the older adult older adults demographic}, in:
  \bibinfo{booktitle}{Artificial Intelligence in Brain and Mental Health:
  Philosophical, Ethical \& Policy Issues}. \bibinfo{publisher}{Springer}, pp.
  \bibinfo{pages}{159--176}.
%Type = Article
\bibitem[{Nguyen et~al.(2021)Nguyen, Raff, Nicholas and
  Holt}]{nguyen2021leveraging}
\bibinfo{author}{Nguyen, A.T.}, \bibinfo{author}{Raff, E.},
  \bibinfo{author}{Nicholas, C.}, \bibinfo{author}{Holt, J.},
  \bibinfo{year}{2021}.
\newblock \bibinfo{title}{Leveraging uncertainty for improved static malware
  detection under extreme false positive constraints}.
\newblock \bibinfo{journal}{arXiv preprint arXiv:2108.04081} .
%Type = Inproceedings
\bibitem[{Nushi et~al.(2018)Nushi, Kamar and Horvitz}]{nushi2018towards}
\bibinfo{author}{Nushi, B.}, \bibinfo{author}{Kamar, E.},
  \bibinfo{author}{Horvitz, E.}, \bibinfo{year}{2018}.
\newblock \bibinfo{title}{Towards accountable ai: Hybrid human-machine analyses
  for characterizing system failure}, in: \bibinfo{booktitle}{Proceedings of
  the AAAI Conference on Human Computation and Crowdsourcing}, pp.
  \bibinfo{pages}{126--135}.
%Type = Article
\bibitem[{Organization et~al.(2016)}]{world2016medicines}
\bibinfo{author}{Organization, W.H.}, et~al., \bibinfo{year}{2016}.
\newblock \bibinfo{title}{Medicines transparency alliance (meta): pathways to
  transparency, accountability an access: cross-case analysis and review of
  phase ii} .
%Type = Article
\bibitem[{Ozga(2020)}]{ozga2020politics}
\bibinfo{author}{Ozga, J.}, \bibinfo{year}{2020}.
\newblock \bibinfo{title}{The politics of accountability}.
\newblock \bibinfo{journal}{Journal of Educational Change}
  \bibinfo{volume}{21}, \bibinfo{pages}{19--35}.
%Type = Article
\bibitem[{Paredes et~al.(2022)Paredes, Teze, Martinez and
  Simari}]{paredes2022heic}
\bibinfo{author}{Paredes, J.N.}, \bibinfo{author}{Teze, J.C.L.},
  \bibinfo{author}{Martinez, M.V.}, \bibinfo{author}{Simari, G.I.},
  \bibinfo{year}{2022}.
\newblock \bibinfo{title}{The heic application framework for implementing
  xai-based socio-technical systems}.
\newblock \bibinfo{journal}{Online Social Networks and Media}
  \bibinfo{volume}{32}, \bibinfo{pages}{100239}.
%Type = Article
\bibitem[{Park et~al.(2021)Park, Hu, Singh, Sylla, Dankwa-Mullan, Koski and
  Das}]{park2021comparison}
\bibinfo{author}{Park, Y.}, \bibinfo{author}{Hu, J.}, \bibinfo{author}{Singh,
  M.}, \bibinfo{author}{Sylla, I.}, \bibinfo{author}{Dankwa-Mullan, I.},
  \bibinfo{author}{Koski, E.}, \bibinfo{author}{Das, A.K.},
  \bibinfo{year}{2021}.
\newblock \bibinfo{title}{Comparison of methods to reduce bias from clinical
  prediction models of postpartum depression}.
\newblock \bibinfo{journal}{JAMA network open} \bibinfo{volume}{4},
  \bibinfo{pages}{e213909--e213909}.
%Type = Article
\bibitem[{Parker(2007)}]{parker2007meta}
\bibinfo{author}{Parker, C.}, \bibinfo{year}{2007}.
\newblock \bibinfo{title}{Meta-regulation: legal accountability for corporate
  social responsibility} .
%Type = Inproceedings
\bibitem[{Pastaltzidis et~al.(2022)Pastaltzidis, Dimitriou, Quezada-Tavarez,
  Aidinlis, Marquenie, Gurzawska and Tzovaras}]{pastaltzidis2022data}
\bibinfo{author}{Pastaltzidis, I.}, \bibinfo{author}{Dimitriou, N.},
  \bibinfo{author}{Quezada-Tavarez, K.}, \bibinfo{author}{Aidinlis, S.},
  \bibinfo{author}{Marquenie, T.}, \bibinfo{author}{Gurzawska, A.},
  \bibinfo{author}{Tzovaras, D.}, \bibinfo{year}{2022}.
\newblock \bibinfo{title}{Data augmentation for fairness-aware machine
  learning: Preventing algorithmic bias in law enforcement systems}, in:
  \bibinfo{booktitle}{2022 ACM Conference on Fairness, Accountability, and
  Transparency}, pp. \bibinfo{pages}{2302--2314}.
%Type = Article
\bibitem[{Pencina et~al.(2012)Pencina, D'Agostino~Sr and
  Demler}]{pencina2012novel}
\bibinfo{author}{Pencina, M.J.}, \bibinfo{author}{D'Agostino~Sr, R.B.},
  \bibinfo{author}{Demler, O.V.}, \bibinfo{year}{2012}.
\newblock \bibinfo{title}{Novel metrics for evaluating improvement in
  discrimination: net reclassification and integrated discrimination
  improvement for normal variables and nested models}.
\newblock \bibinfo{journal}{Statistics in medicine} \bibinfo{volume}{31},
  \bibinfo{pages}{101--113}.
%Type = Article
\bibitem[{Pirraglia and Kravitz(2013)}]{pirraglia2013social}
\bibinfo{author}{Pirraglia, P.A.}, \bibinfo{author}{Kravitz, R.L.},
  \bibinfo{year}{2013}.
\newblock \bibinfo{title}{Social media: new opportunities, new ethical
  concerns}.
\newblock \bibinfo{journal}{Journal of general internal medicine}
  \bibinfo{volume}{28}, \bibinfo{pages}{165--166}.
%Type = Inproceedings
\bibitem[{Raji et~al.(2020)Raji, Smart, White, Mitchell, Gebru, Hutchinson,
  Smith-Loud, Theron and Barnes}]{raji2020closing}
\bibinfo{author}{Raji, I.D.}, \bibinfo{author}{Smart, A.},
  \bibinfo{author}{White, R.N.}, \bibinfo{author}{Mitchell, M.},
  \bibinfo{author}{Gebru, T.}, \bibinfo{author}{Hutchinson, B.},
  \bibinfo{author}{Smith-Loud, J.}, \bibinfo{author}{Theron, D.},
  \bibinfo{author}{Barnes, P.}, \bibinfo{year}{2020}.
\newblock \bibinfo{title}{Closing the ai accountability gap: Defining an
  end-to-end framework for internal algorithmic auditing}, in:
  \bibinfo{booktitle}{Proceedings of the 2020 conference on fairness,
  accountability, and transparency}, pp. \bibinfo{pages}{33--44}.
%Type = Article
\bibitem[{Reich(2018)}]{reich2018core}
\bibinfo{author}{Reich, M.R.}, \bibinfo{year}{2018}.
\newblock \bibinfo{title}{The core roles of transparency and accountability in
  the governance of global health public--private partnerships}.
\newblock \bibinfo{journal}{Health Systems \& Reform} \bibinfo{volume}{4},
  \bibinfo{pages}{239--248}.
%Type = Article
\bibitem[{Rosenfeld and Richardson(2019)}]{rosenfeld2019explainability}
\bibinfo{author}{Rosenfeld, A.}, \bibinfo{author}{Richardson, A.},
  \bibinfo{year}{2019}.
\newblock \bibinfo{title}{Explainability in human--agent systems}.
\newblock \bibinfo{journal}{Autonomous Agents and Multi-Agent Systems}
  \bibinfo{volume}{33}, \bibinfo{pages}{673--705}.
%Type = Inproceedings
\bibitem[{Saha et~al.(2020)Saha, Schumann, Mcelfresh, Dickerson, Mazurek and
  Tschantz}]{saha2020measuring}
\bibinfo{author}{Saha, D.}, \bibinfo{author}{Schumann, C.},
  \bibinfo{author}{Mcelfresh, D.}, \bibinfo{author}{Dickerson, J.},
  \bibinfo{author}{Mazurek, M.}, \bibinfo{author}{Tschantz, M.},
  \bibinfo{year}{2020}.
\newblock \bibinfo{title}{Measuring non-expert comprehension of machine
  learning fairness metrics}, in: \bibinfo{booktitle}{International Conference
  on Machine Learning}, \bibinfo{organization}{PMLR}. pp.
  \bibinfo{pages}{8377--8387}.
%Type = Article
\bibitem[{Saleiro et~al.(2018)Saleiro, Kuester, Hinkson, London, Stevens,
  Anisfeld, Rodolfa and Ghani}]{saleiro2018aequitas}
\bibinfo{author}{Saleiro, P.}, \bibinfo{author}{Kuester, B.},
  \bibinfo{author}{Hinkson, L.}, \bibinfo{author}{London, J.},
  \bibinfo{author}{Stevens, A.}, \bibinfo{author}{Anisfeld, A.},
  \bibinfo{author}{Rodolfa, K.T.}, \bibinfo{author}{Ghani, R.},
  \bibinfo{year}{2018}.
\newblock \bibinfo{title}{Aequitas: A bias and fairness audit toolkit}.
\newblock \bibinfo{journal}{arXiv preprint arXiv:1811.05577} .
%Type = Article
\bibitem[{Sanfey(2007)}]{sanfey2007social}
\bibinfo{author}{Sanfey, A.G.}, \bibinfo{year}{2007}.
\newblock \bibinfo{title}{Social decision-making: insights from game theory and
  neuroscience}.
\newblock \bibinfo{journal}{Science} \bibinfo{volume}{318},
  \bibinfo{pages}{598--602}.
%Type = Inproceedings
\bibitem[{Saxena et~al.(2019)Saxena, Huang, DeFilippis, Radanovic, Parkes and
  Liu}]{saxena2019fairness}
\bibinfo{author}{Saxena, N.A.}, \bibinfo{author}{Huang, K.},
  \bibinfo{author}{DeFilippis, E.}, \bibinfo{author}{Radanovic, G.},
  \bibinfo{author}{Parkes, D.C.}, \bibinfo{author}{Liu, Y.},
  \bibinfo{year}{2019}.
\newblock \bibinfo{title}{How do fairness definitions fare? examining public
  attitudes towards algorithmic definitions of fairness}, in:
  \bibinfo{booktitle}{Proceedings of the 2019 AAAI/ACM Conference on AI,
  Ethics, and Society}, pp. \bibinfo{pages}{99--106}.
%Type = Book
\bibitem[{Sharma(2019)}]{sharma2019data}
\bibinfo{author}{Sharma, S.}, \bibinfo{year}{2019}.
\newblock \bibinfo{title}{Data privacy and GDPR handbook}.
\newblock \bibinfo{publisher}{John Wiley \& Sons}.
%Type = Article
\bibitem[{Shneiderman(2020)}]{shneiderman2020bridging}
\bibinfo{author}{Shneiderman, B.}, \bibinfo{year}{2020}.
\newblock \bibinfo{title}{Bridging the gap between ethics and practice:
  guidelines for reliable, safe, and trustworthy human-centered ai systems}.
\newblock \bibinfo{journal}{ACM Transactions on Interactive Intelligent Systems
  (TiiS)} \bibinfo{volume}{10}, \bibinfo{pages}{1--31}.
%Type = Article
\bibitem[{Singhal et~al.(2022)Singhal, Baxi, Mago et~al.}]{singhal2022synergy}
\bibinfo{author}{Singhal, A.}, \bibinfo{author}{Baxi, M.K.},
  \bibinfo{author}{Mago, V.}, et~al., \bibinfo{year}{2022}.
\newblock \bibinfo{title}{Synergy between public and private health care
  organizations during covid-19 on twitter: Sentiment and engagement analysis
  using forecasting models}.
\newblock \bibinfo{journal}{JMIR Medical Informatics} \bibinfo{volume}{10},
  \bibinfo{pages}{e37829}.
%Type = Article
\bibitem[{Slack et~al.(2019)Slack, Friedler, Scheidegger and
  Roy}]{slack2019assessing}
\bibinfo{author}{Slack, D.}, \bibinfo{author}{Friedler, S.A.},
  \bibinfo{author}{Scheidegger, C.}, \bibinfo{author}{Roy, C.D.},
  \bibinfo{year}{2019}.
\newblock \bibinfo{title}{Assessing the local interpretability of machine
  learning models}.
\newblock \bibinfo{journal}{arXiv preprint arXiv:1902.03501} .
%Type = Article
\bibitem[{Sokol and Flach(2019)}]{sokol2019counterfactual}
\bibinfo{author}{Sokol, K.}, \bibinfo{author}{Flach, P.A.},
  \bibinfo{year}{2019}.
\newblock \bibinfo{title}{Counterfactual explanations of machine learning
  predictions: Opportunities and challenges for ai safety.}
\newblock \bibinfo{journal}{SafeAI@ AAAI} .
%Type = Article
\bibitem[{Someh et~al.(2019)Someh, Davern, Breidbach and
  Shanks}]{someh2019ethical}
\bibinfo{author}{Someh, I.}, \bibinfo{author}{Davern, M.},
  \bibinfo{author}{Breidbach, C.F.}, \bibinfo{author}{Shanks, G.},
  \bibinfo{year}{2019}.
\newblock \bibinfo{title}{Ethical issues in big data analytics: A stakeholder
  perspective}.
\newblock \bibinfo{journal}{Communications of the Association for Information
  Systems} \bibinfo{volume}{44}, \bibinfo{pages}{34}.
%Type = Article
\bibitem[{S{\o}rensen(2018)}]{sorensen2018health}
\bibinfo{author}{S{\o}rensen, K.}, \bibinfo{year}{2018}.
\newblock \bibinfo{title}{Health literacy: A key attribute for urban settings}.
\newblock \bibinfo{journal}{Optimizing Health Literacy for Improved Clinical
  Practices} , \bibinfo{pages}{1--16}.
%Type = Article
\bibitem[{Stanberry(2006)}]{stanberry2006legal}
\bibinfo{author}{Stanberry, B.}, \bibinfo{year}{2006}.
\newblock \bibinfo{title}{Legal and ethical aspects of telemedicine}.
\newblock \bibinfo{journal}{Journal of telemedicine and telecare}
  \bibinfo{volume}{12}, \bibinfo{pages}{166--175}.
%Type = Article
\bibitem[{Stellefson et~al.(2020)Stellefson, Paige, Chaney and
  Chaney}]{stellefson2020evolving}
\bibinfo{author}{Stellefson, M.}, \bibinfo{author}{Paige, S.R.},
  \bibinfo{author}{Chaney, B.H.}, \bibinfo{author}{Chaney, J.D.},
  \bibinfo{year}{2020}.
\newblock \bibinfo{title}{Evolving role of social media in health promotion:
  updated responsibilities for health education specialists}.
\newblock \bibinfo{journal}{International journal of environmental research and
  public health} \bibinfo{volume}{17}, \bibinfo{pages}{1153}.
%Type = Article
\bibitem[{Stepin et~al.(2021)Stepin, Alonso, Catala and
  Pereira-Fari{\~n}a}]{stepin2021survey}
\bibinfo{author}{Stepin, I.}, \bibinfo{author}{Alonso, J.M.},
  \bibinfo{author}{Catala, A.}, \bibinfo{author}{Pereira-Fari{\~n}a, M.},
  \bibinfo{year}{2021}.
\newblock \bibinfo{title}{A survey of contrastive and counterfactual
  explanation generation methods for explainable artificial intelligence}.
\newblock \bibinfo{journal}{IEEE Access} \bibinfo{volume}{9},
  \bibinfo{pages}{11974--12001}.
%Type = Article
\bibitem[{Swan(2009)}]{swan2009emerging}
\bibinfo{author}{Swan, M.}, \bibinfo{year}{2009}.
\newblock \bibinfo{title}{Emerging patient-driven health care models: an
  examination of health social networks, consumer personalized medicine and
  quantified self-tracking}.
\newblock \bibinfo{journal}{International journal of environmental research and
  public health} \bibinfo{volume}{6}, \bibinfo{pages}{492--525}.
%Type = Inproceedings
\bibitem[{Tao et~al.(2022)Tao, Sun, Han, Fang and Zhang}]{tao2022ruler}
\bibinfo{author}{Tao, G.}, \bibinfo{author}{Sun, W.}, \bibinfo{author}{Han,
  T.}, \bibinfo{author}{Fang, C.}, \bibinfo{author}{Zhang, X.},
  \bibinfo{year}{2022}.
\newblock \bibinfo{title}{Ruler: discriminative and iterative adversarial
  training for deep neural network fairness}, in:
  \bibinfo{booktitle}{Proceedings of the 30th ACM Joint European Software
  Engineering Conference and Symposium on the Foundations of Software
  Engineering}, pp. \bibinfo{pages}{1173--1184}.
%Type = Article
\bibitem[{Umbrello and Van~de Poel(2021)}]{umbrello2021mapping}
\bibinfo{author}{Umbrello, S.}, \bibinfo{author}{Van~de Poel, I.},
  \bibinfo{year}{2021}.
\newblock \bibinfo{title}{Mapping value sensitive design onto ai for social
  good principles}.
\newblock \bibinfo{journal}{AI and Ethics} \bibinfo{volume}{1},
  \bibinfo{pages}{283--296}.
%Type = Incollection
\bibitem[{Unerman(2010)}]{unerman2010stakeholder}
\bibinfo{author}{Unerman, J.}, \bibinfo{year}{2010}.
\newblock \bibinfo{title}{Stakeholder engagement and dialogue}, in:
  \bibinfo{booktitle}{Sustainability accounting and accountability}.
  \bibinfo{publisher}{Routledge}, pp. \bibinfo{pages}{105--122}.
%Type = Article
\bibitem[{Valko and Hauskrecht(2010)}]{valko2010feature}
\bibinfo{author}{Valko, M.}, \bibinfo{author}{Hauskrecht, M.},
  \bibinfo{year}{2010}.
\newblock \bibinfo{title}{Feature importance analysis for patient management
  decisions}.
\newblock \bibinfo{journal}{Studies in health technology and informatics}
  \bibinfo{volume}{160}, \bibinfo{pages}{861}.
%Type = Article
\bibitem[{Ventola(2014)}]{ventola2014social}
\bibinfo{author}{Ventola, C.L.}, \bibinfo{year}{2014}.
\newblock \bibinfo{title}{Social media and health care professionals: benefits,
  risks, and best practices}.
\newblock \bibinfo{journal}{Pharmacy and therapeutics} \bibinfo{volume}{39},
  \bibinfo{pages}{491}.
%Type = Article
\bibitem[{Vergeer et~al.(2021)Vergeer, van Schaik and
  Sjerps}]{vergeer2021measuring}
\bibinfo{author}{Vergeer, P.}, \bibinfo{author}{van Schaik, Y.},
  \bibinfo{author}{Sjerps, M.}, \bibinfo{year}{2021}.
\newblock \bibinfo{title}{Measuring calibration of likelihood-ratio systems: a
  comparison of four metrics, including a new metric devpav}.
\newblock \bibinfo{journal}{Forensic Science International}
  \bibinfo{volume}{321}, \bibinfo{pages}{110722}.
%Type = Article
\bibitem[{Vesnic-Alujevic et~al.(2020)Vesnic-Alujevic, Nascimento and
  Polvora}]{vesnic2020societal}
\bibinfo{author}{Vesnic-Alujevic, L.}, \bibinfo{author}{Nascimento, S.},
  \bibinfo{author}{Polvora, A.}, \bibinfo{year}{2020}.
\newblock \bibinfo{title}{Societal and ethical impacts of artificial
  intelligence: Critical notes on european policy frameworks}.
\newblock \bibinfo{journal}{Telecommunications Policy} \bibinfo{volume}{44},
  \bibinfo{pages}{101961}.
%Type = Article
\bibitem[{Wachter et~al.(2017)Wachter, Mittelstadt and
  Floridi}]{wachter2017transparent}
\bibinfo{author}{Wachter, S.}, \bibinfo{author}{Mittelstadt, B.},
  \bibinfo{author}{Floridi, L.}, \bibinfo{year}{2017}.
\newblock \bibinfo{title}{Transparent, explainable, and accountable ai for
  robotics}.
\newblock \bibinfo{journal}{Science robotics} \bibinfo{volume}{2},
  \bibinfo{pages}{eaan6080}.
%Type = Article
\bibitem[{Weiskopf et~al.(2013)Weiskopf, Hripcsak, Swaminathan and
  Weng}]{weiskopf2013defining}
\bibinfo{author}{Weiskopf, N.G.}, \bibinfo{author}{Hripcsak, G.},
  \bibinfo{author}{Swaminathan, S.}, \bibinfo{author}{Weng, C.},
  \bibinfo{year}{2013}.
\newblock \bibinfo{title}{Defining and measuring completeness of electronic
  health records for secondary use}.
\newblock \bibinfo{journal}{Journal of biomedical informatics}
  \bibinfo{volume}{46}, \bibinfo{pages}{830--836}.
%Type = Article
\bibitem[{Weiss et~al.(2018)Weiss, Nelson, Gibson, Temperley, Peedell, Lieber,
  Hancher, Poyart, Belchior, Fullman et~al.}]{weiss2018global}
\bibinfo{author}{Weiss, D.J.}, \bibinfo{author}{Nelson, A.},
  \bibinfo{author}{Gibson, H.}, \bibinfo{author}{Temperley, W.},
  \bibinfo{author}{Peedell, S.}, \bibinfo{author}{Lieber, A.},
  \bibinfo{author}{Hancher, M.}, \bibinfo{author}{Poyart, E.},
  \bibinfo{author}{Belchior, S.}, \bibinfo{author}{Fullman, N.}, et~al.,
  \bibinfo{year}{2018}.
\newblock \bibinfo{title}{A global map of travel time to cities to assess
  inequalities in accessibility in 2015}.
\newblock \bibinfo{journal}{Nature} \bibinfo{volume}{553},
  \bibinfo{pages}{333--336}.
%Type = Article
\bibitem[{Weng et~al.(2018)Weng, Zhang, Chen, Yi, Su, Gao, Hsieh and
  Daniel}]{weng2018evaluating}
\bibinfo{author}{Weng, T.W.}, \bibinfo{author}{Zhang, H.},
  \bibinfo{author}{Chen, P.Y.}, \bibinfo{author}{Yi, J.}, \bibinfo{author}{Su,
  D.}, \bibinfo{author}{Gao, Y.}, \bibinfo{author}{Hsieh, C.J.},
  \bibinfo{author}{Daniel, L.}, \bibinfo{year}{2018}.
\newblock \bibinfo{title}{Evaluating the robustness of neural networks: An
  extreme value theory approach}.
\newblock \bibinfo{journal}{arXiv preprint arXiv:1801.10578} .
%Type = Inproceedings
\bibitem[{Wieringa(2020)}]{wieringa2020account}
\bibinfo{author}{Wieringa, M.}, \bibinfo{year}{2020}.
\newblock \bibinfo{title}{What to account for when accounting for algorithms},
  in: \bibinfo{booktitle}{Proceedings of the 2020 Conference on Fairness,
  Accountability, and Transparency}, pp. \bibinfo{pages}{1--18}.
%Type = Article
\bibitem[{Wright(2011)}]{wright2011framework}
\bibinfo{author}{Wright, D.}, \bibinfo{year}{2011}.
\newblock \bibinfo{title}{A framework for the ethical impact assessment of
  information technology}.
\newblock \bibinfo{journal}{Ethics and information technology}
  \bibinfo{volume}{13}, \bibinfo{pages}{199--226}.
%Type = Article
\bibitem[{Xiong et~al.(2020)Xiong, Cui, Liu, Zhao, Hu and
  Hu}]{xiong2020evaluating}
\bibinfo{author}{Xiong, Z.}, \bibinfo{author}{Cui, Y.}, \bibinfo{author}{Liu,
  Z.}, \bibinfo{author}{Zhao, Y.}, \bibinfo{author}{Hu, M.},
  \bibinfo{author}{Hu, J.}, \bibinfo{year}{2020}.
\newblock \bibinfo{title}{Evaluating explorative prediction power of machine
  learning algorithms for materials discovery using k-fold forward
  cross-validation}.
\newblock \bibinfo{journal}{Computational Materials Science}
  \bibinfo{volume}{171}, \bibinfo{pages}{109203}.
%Type = Article
\bibitem[{Xu et~al.(2022)Xu, Xiao, Wang, Ning, Shenkman, Bian and
  Wang}]{xu2022algorithmic}
\bibinfo{author}{Xu, J.}, \bibinfo{author}{Xiao, Y.}, \bibinfo{author}{Wang,
  W.H.}, \bibinfo{author}{Ning, Y.}, \bibinfo{author}{Shenkman, E.A.},
  \bibinfo{author}{Bian, J.}, \bibinfo{author}{Wang, F.}, \bibinfo{year}{2022}.
\newblock \bibinfo{title}{Algorithmic fairness in computational medicine}.
\newblock \bibinfo{journal}{EBioMedicine} \bibinfo{volume}{84},
  \bibinfo{pages}{104250}.
%Type = Article
\bibitem[{Yao et~al.(2021)Yao, Chen, Ye, Jin and Ren}]{yao2021refining}
\bibinfo{author}{Yao, H.}, \bibinfo{author}{Chen, Y.}, \bibinfo{author}{Ye,
  Q.}, \bibinfo{author}{Jin, X.}, \bibinfo{author}{Ren, X.},
  \bibinfo{year}{2021}.
\newblock \bibinfo{title}{Refining language models with compositional
  explanations}.
\newblock \bibinfo{journal}{Advances in Neural Information Processing Systems}
  \bibinfo{volume}{34}, \bibinfo{pages}{8954--8967}.
%Type = Article
\bibitem[{Yao and Huang(2017)}]{yao2017beyond}
\bibinfo{author}{Yao, S.}, \bibinfo{author}{Huang, B.}, \bibinfo{year}{2017}.
\newblock \bibinfo{title}{Beyond parity: Fairness objectives for collaborative
  filtering}.
\newblock \bibinfo{journal}{Advances in neural information processing systems}
  \bibinfo{volume}{30}.
%Type = Article
\bibitem[{Zafar et~al.(2019)Zafar, Valera, Gomez-Rodriguez and
  Gummadi}]{zafar2019fairness}
\bibinfo{author}{Zafar, M.B.}, \bibinfo{author}{Valera, I.},
  \bibinfo{author}{Gomez-Rodriguez, M.}, \bibinfo{author}{Gummadi, K.P.},
  \bibinfo{year}{2019}.
\newblock \bibinfo{title}{Fairness constraints: A flexible approach for fair
  classification}.
\newblock \bibinfo{journal}{The Journal of Machine Learning Research}
  \bibinfo{volume}{20}, \bibinfo{pages}{2737--2778}.
%Type = Article
\bibitem[{Zafar and Khan(2021)}]{zafar2021deterministic}
\bibinfo{author}{Zafar, M.R.}, \bibinfo{author}{Khan, N.},
  \bibinfo{year}{2021}.
\newblock \bibinfo{title}{Deterministic local interpretable model-agnostic
  explanations for stable explainability}.
\newblock \bibinfo{journal}{Machine Learning and Knowledge Extraction}
  \bibinfo{volume}{3}, \bibinfo{pages}{525--541}.
%Type = Article
\bibitem[{Zaki et~al.(2021)Zaki, Jena and Chandra}]{zaki2021supporting}
\bibinfo{author}{Zaki, M.M.}, \bibinfo{author}{Jena, A.B.},
  \bibinfo{author}{Chandra, A.}, \bibinfo{year}{2021}.
\newblock \bibinfo{title}{Supporting value-based health care-aligning financial
  and legal accountability}.
\newblock \bibinfo{journal}{The New England journal of medicine}
  \bibinfo{volume}{385}, \bibinfo{pages}{965--967}.
%Type = Inproceedings
\bibitem[{Zhai et~al.(2015)Zhai, Cohen and Lafferty}]{zhai2015beyond}
\bibinfo{author}{Zhai, C.}, \bibinfo{author}{Cohen, W.W.},
  \bibinfo{author}{Lafferty, J.}, \bibinfo{year}{2015}.
\newblock \bibinfo{title}{Beyond independent relevance: methods and evaluation
  metrics for subtopic retrieval}, in: \bibinfo{booktitle}{Acm sigir forum},
  \bibinfo{organization}{ACM New York, NY, USA}. pp. \bibinfo{pages}{2--9}.
%Type = Article
\bibitem[{Zhang and Zhou(2019)}]{zhang2019fairness}
\bibinfo{author}{Zhang, Y.}, \bibinfo{author}{Zhou, L.}, \bibinfo{year}{2019}.
\newblock \bibinfo{title}{Fairness assessment for artificial intelligence in
  financial industry}.
\newblock \bibinfo{journal}{arXiv preprint arXiv:1912.07211} .
%Type = Article
\bibitem[{Zhao et~al.(2020)Zhao, Lovreglio and Nilsson}]{zhao2020modelling}
\bibinfo{author}{Zhao, X.}, \bibinfo{author}{Lovreglio, R.},
  \bibinfo{author}{Nilsson, D.}, \bibinfo{year}{2020}.
\newblock \bibinfo{title}{Modelling and interpreting pre-evacuation
  decision-making using machine learning}.
\newblock \bibinfo{journal}{Automation in Construction} \bibinfo{volume}{113},
  \bibinfo{pages}{103140}.

\end{thebibliography}

%\vskip3pt

\end{document}